\documentclass[twocolumn,aps,showpacs,preprintnumbers,amsmath,amssymb,prb]{revtex4}
\usepackage{graphicx}
\usepackage{dcolumn}
\usepackage{bm}
\usepackage[sort&compress]{natbib}
\usepackage{subfigure}

\begin{document}

\title{From vortex molecules to the Abrikosov lattice in thin mesoscopic superconducting disks}
\author{L. R. E. Cabral}
\email{leonardo.cabral@ua.ac.be} \affiliation{Departement
Natuurkunde,
Universiteit Antwerpen (Campus Drie Eiken), \\
Universiteitsplein 1, B-2610 Antwerpen, Belgium}
\author{B. J. Baelus}
\email{ben.baelus@ua.ac.be} \altaffiliation[Present address:
]{Institute of Materials Science, University of Tsukuba, Tsukuba
305-8573, Japan}\affiliation{Departement Natuurkunde,
Universiteit Antwerpen (Campus Drie Eiken), \\
Universiteitsplein 1, B-2610 Antwerpen, Belgium}
\author{F. M. Peeters}
\email{francois.peeters@ua.ac.be} \affiliation{Departement
Natuurkunde,
Universiteit Antwerpen (Campus Drie Eiken), \\
Universiteitsplein 1, B-2610 Antwerpen, Belgium}
\date{\today}

\begin{abstract}
Stable vortex states are studied in large superconducting thin
disks (for numerical purposes we considered with radius $R = 50
\xi$). Configurations containing more than $700$ vortices were
obtained using two different approaches: the nonlinear
Ginzburg-Landau (GL) theory and the London approximation. To
obtain better agreement with results from the GL theory  we
generalized the London theory by including the spatial variation
of the order parameter following Clem's ansatz. We find that
configurations calculated in the London limit are also stable
within the Ginzburg-Landau theory for up to $\sim 230$ vortices.
For large values of the vorticity (typically, $L \gtrsim 100$),
the vortices are arranged in an Abrikosov lattice in the center of
the disk, which is surrounded by at least two circular shells of
vortices. A Voronoi construction is used to identify the defects
present in  the ground state vortex configurations. Such defects
cluster near the edge of the disk, but for large $L$ also grain
boundaries are found which extend up to the center of the disk.
\end{abstract}


\maketitle

\section{Introduction}

Vortices appear in several branches of physics, such as fluid
dynamics,~\cite{LandauFluid} superfluidity,~\cite{Tilley}
Bose-Einstein (B-E)
condensates,~\cite{PRL83_2498,RMP71_463,JCMP13_135} and
superconductivity.~\cite{Abrikosov_57,Abrikosov} The vortex is
usually described by a field (for instance, the velocity field)
which diverges as $r^{-1}$ as one approaches its
core.~\cite{PhysC369_1} They can be treated as quasiparticles,
since they can be created or destroyed, they interact with each
other and with the interfaces. Unlike in fluid dynamics, in
superfluids (including here superconductors and B-E condensates)
vortices are quantized objects. In superconductors, for example,
they carry a magnetic flux which is a multiple of the flux
quantum, $\Phi_0 = hc / 2e$, and are characterized by a core of
area $\xi^2$ -- where the superconductivity is highly depreciated
-- surrounded by superconducting currents (screened at distances
of order $\lambda$). Here, $\xi$ is the coherence length. They
have been intensively studied, since Abrikosov~\cite{Abrikosov_57}
predicted their existence from the solution of the Ginzburg-Landau
(GL) equations in a type-II superconductor for $H_{c1} < H <
H_{c2}$ (see also Refs.~\onlinecite{Brandte} and
~\onlinecite{BrandtPRB68_4506}). In an infinite, and defect free
superconductor, vortices arrange themselves in an hexagonal
(Abrikosov) lattice.

A detailed phenomenological description of the superconducting
state can be derived from the GL theory,~\cite{G-L_50} by means of
two parameters: the complex order parameter, $\Psi$, which is
related to the superconducting electron density, and the vector
potential, $\bm{A}$. For $H_{c1} \le H \ll H_{c2}$, each vortex
can be viewed as a particle, since inter-vortex separations, $a$,
are such that $\xi \ll a \sim \lambda$ -- assuring that vortex
cores do not overlap -- and the major role between vortex-vortex
interactions is played by the superconducting shielding currents.
In such cases the London limit turns out to be a good
approximation of the GL theory, becoming better for higher values
of $\kappa$ (see for example
Refs.~\onlinecite{PR147_140,PR159_330,deGennesbook,Abrikosov}). In
this approximation, the superconducting electron density is
considered constant throughout the entire superconductor and the
vortex cores are represented by singularities in the phase of the
order parameter. This allows to treat vortices as particles.

In a thin film of thickness $d$, the effective magnetic field
shielding length turns out to be the effective penetration depth,
$\Lambda = \lambda^2 / d$, instead of $\lambda$.~\cite{APL5_65} At
distances $r \ll \Lambda$ the electromagnetic interaction is still
logarithmic, as in the three dimensional case, but with screening
length $\Lambda$ [However the perpendicular magnetic field and the
shielding currents decay as $r^{-3}$ and $r^{-2}$ far away from
the vortex core for $r \ll \Lambda$, instead as $\exp(-r/\lambda)$
in the bulk case.] Similarly as in the bulk case, in a thin film
vortices also form an hexagonal Abrikosov
lattice.~\cite{PR159_330}

In mesoscopic superconductors both the geometry and size of the
specimen influence the vortex configurations, due to the
interaction between vortices and the surface. Therefore, for small
enough samples (with sizes comparable to $\xi$), the conventional
hexagonal lattice predicted by Abrikosov no longer exists, and
vortex configurations adjust to the sample geometry, yielding some
kind of vortex molecule
states.~\cite{PRB65_104515,PRB65_140503,PPRB58,PRB63,Baelus_PRB03}
For example, vortices arrange themselves in ringlike structures in
disks with radii ($R$) a few times
$\xi$.~\cite{GN390,PRL79d,PRB57v,PRL81v,PRL83v,
PRB59d,SM25d,PRB62s,GN407,PPRL84,PPRB58,PRB63,Baelus_PRB03} Such
patterns show similarities to what is observed in electrons in
artificial atoms, where particles obey specific rules for shell
filling and exhibit magic numbers. Nevertheless when overlapping
of vortices starts to take place, discrepancies between vortices
and a picture based on particles arise, such as the formation of
giant vortex states. Also, vortex-antivortex configurations may
become possible for non circular
geometries.~\cite{PRB65_012509,V-AVall,PRB67_134527}

Within the London limit the vortex interaction potential in a thin
disk of arbitrary radius was first calculated by
Fetter~\cite{PRB22_1200}. Also in the London limit, vortex
configurations up to $L = 8$ were studied by Buzdin and
Brison~\cite{Buzdin} for $\Lambda \gg R$ (where demagnetization
effects can be neglected). In the latter limit it is possible to
substitute the interaction between the vortices and the disk
border by the interaction between vortices and their images (see
also Ref.~\onlinecite{BuzDaum}). Within the London limit one is
able to find analytical expressions for the energy and forces of
an arbitrary arrangement of vortices inside the disk, since
vortices can be treated as particles. Vortices considered as
particles were also studied by Monte Carlo and Molecular Dynamics
simulations. In Ref.~\onlinecite{PhysC183_212} vortex
configurations with up to 2000 vortices were studied and an
hexagonal lattice was found for thin disks, although they did not
consider the vortex interaction with the disk edge. Vortex
molecules in long cylinders with radius much larger than $\lambda$
were studied by Venegas and Sardella.~\cite{VenegasSardella} Other
geometries were investigated in
Refs.~\cite{clecioleo,sardellaetal_00}, for example.

In this paper we will study multivortex states where many vortices
nucleate, yielding a triangular lattice in the center of the disk
and a ringlike structure close to the edges. Within the GL
framework several other works have been reported regarding vortex
states in thin disks,~\cite{GN390,PRL79d,PRB57v,PRL81v,PRL83v,
PRB59d,SM25d,PRB62s,GN407,PPRL84,PPRB58,PRB63} but they were
limited to much smaller disk radius. In such small systems the
formation of multivortex states with high vorticity is not allowed
and, consequently, it was not possible to study the transition
from a ringlike structure to an Abrikosov lattice, which is the
subject of the present paper.

This paper is organized as follows. The theoretical approach is
described in Section~\ref{secii}. In Section~\ref{seciii} low
vorticity states obtained within the GL and the London frameworks
are compared. In Sections~\ref{seciv} and~\ref{secv}
configurations with up to 700 vortices are investigated,
respectively, by showing the existence of an Abrikosov lattice in
the center of the disk and by examining the role of topological
defects in the lattice in order to adjust the hexagonal lattice to
the radial symmetry close to the disk edge. Surface
superconductivity in the $R = 50\xi$ disk is briefly analyzed in
Section~\ref{secvi}. Our conclusions are given in
Section~\ref{secviii}.

\section{Theoretical approach}
\label{secii}

 For our numerical calculation we used a thin disk of
radius $R = 50 \xi$ and thickness $d$, in which $\Lambda =
\lambda^2 / d \gg \xi \gg d$, surrounded by vacuum and in the
presence of a uniform perpendicular magnetic field $\bm{H}_{0}$.
In this regime, the demagnetization effects can be neglected,
allowing one to assume $\bm{H} \approx \bm{H}_{0}$. Vortex states
in mesoscopic thin disks were investigated by us using both the
Ginzburg-Landau (GL) theory and the London approximation with the
London gauge $\nabla\cdot\bm{A}=0$. Dimensionless variables are
used, i.e., the distance is measured in units of the coherence
length $\xi $, the vector potential in $c\hbar/2e\xi $ and the
magnetic field in $H_{c2}=c\hbar/2e\xi ^{2}=\kappa \sqrt{2}H_{c}$.
The average energy density is written in units of $H_c^2 / 8\pi$
(we shall refer to it as simply the energy of the system). Also,
the vorticity or the number of vortices in the system will be
denoted by $L$ (an analogue to the total angular momentum).
\cite{PRL81v,Baelus_PRB03} Moreover, whenever the distinction
among different configurations with the same $L$ would be
necessary, we use the notation presented in
Ref.~\onlinecite{Baelus_PRB03} to denote the vortex
configurations, e.g., for $L = 6$, $(1,5)$ means $1$ vortex in the
center with 5 around it, and $(6)$ represents $6$ vortices with
none of them in the center of the disk.

In the framework of the GL theory, the GL equations are solved
numerically according to the approach of Schweigert and Peeters.
\cite{PRB57v,PRL81v} As we are in the limit $(d\ll \xi ,\lambda
)$, the Ginzburg-Landau equations can be averaged over the disk
thickness, leading to the following system of equations,
\begin{equation}
\left( -i{\nabla }_{2D}-\bm{A}\right) ^{2}\Psi =\Psi \left(
1-\left| \Psi \right| ^{2}\right) \label{lijn1}
\end{equation}
and
\begin{equation}
-\Delta _{3D}\bm{A}= \bm{\jmath}    \label{lijn2},
\end{equation}
where the supercurrent density is defined by the following,
\begin{eqnarray}
\frac{\kappa ^{2}}{d}\bm{\jmath} & = & \delta \left( z\right)
\left[\frac{1}{2i}\left( \Psi ^{\ast } \nabla_{2D}\Psi -\Psi
\nabla_{2D} \Psi ^{\ast}\right) - \left| \Psi \right|
^{2}\bm{A}\right ] \nonumber\\
& = & \delta \left( z\right) \left| \Psi \right|
^{2}\left(\nabla_{2D} \theta - \bm{A} \right) = \delta \left(
z\right) \left| \Psi \right| ^{2}\bm{\Pi}\label{lijn3}.
\end{eqnarray}
Above, the superconducting wavefunction, $\Psi = |\Psi|
e^{i\theta}$, satisfies the boundary conditions $\left. \left(
-i\nabla_{2D}-\bm{A}\right) \Psi \right| _{n}=0$ normal to the
sample surface and $\bm{A}=\bm{A}_0 = \frac{1}{2}H_{0}\rho
\hat\phi$ (since demagnetization effects can be neglected). Here
$\hat \phi$ is the unit vector in the azimuthal direction. The
indices 2D, 3D refer to two- and three-dimensional operators,
respectively. The dimensionless GL energy density is given by
\begin{subequations}
\label{GLen}
\begin{eqnarray}
\mathcal{G}  =  \mathcal{G}_{\rm core} +  \mathcal{G}_{\rm em},
\label{GLen1}
\end{eqnarray}
where
\begin{eqnarray}
 \!\!\!\!\mathcal{G}_{\rm core}
 &\! = \!& \frac{1}{V}\int_{V}\left[-2 \left| \Psi \right| ^{2} + \left|
\Psi \right| ^{4} + 2\left(\nabla_{2D} \left| \Psi
\right|\right)^{2}\right]\! dV , \label{GLcore}\\
\!\!\!\!\mathcal{G}_{\rm em}
 &\! =\! & \frac{1}{V}\int_{V}\left[2 \left| \Psi \right| ^{2}\Pi^2 +
 2\kappa^2\left(\bm{H} - \bm{H}_0\right)^2\right] dV, \label{GLem}
\end{eqnarray}
\end{subequations}
are the core and the electromagnetic energies, respectively, and
the integrations are to be performed over the sample volume $V$.
As demagnetization effects can be disregarded, the above equation
reduces to
\begin{equation}
\mathcal{G} =-\frac{1}{V}\int_{V}\left| \Psi \right| ^{4} dV,
\label{GLPsi4}
\end{equation}
which was actually the expression used to compute the energy of
the vortex configurations within the GL theory. For now on the
symbol $\nabla$ will be used for the two-dimensional gradient
operator.

The system of Eqs.~(\ref{lijn1},\ref{lijn2}) were solved by using
the approach of Ref.~\cite{PRL81v} for circular disks. A
finite-difference representation for the order parameter is used
on an uniform 2D square grid (x,y), with typically $512\times 512$
grid points for the area of the superconductor, which allows to
have at least 5 grid points inside a length of the order of $\xi$.
We also use the link variable approach, \cite{Kato} and an
iteration procedure based on the Gauss-Seidel technique to find
$\Psi$. Starting from different randomly generated initial
conditions and at some specified magnetic field, the steady-state
solutions of Eqs.~(\ref{lijn1},\ref{lijn2}) yield different vortex
configurations, either stable or meta-stable states.

In the London approximation, the order parameter is considered
uniform throughout the disk, except for small regions with areas
of the order of $\xi^2$, where it drops to zero. This can only be
accomplished when $\kappa \gg 1$. Then the energy of the system is
purely electromagnetic and it is given by the sum of the
supercurrent and the magnetic field energies
\begin{eqnarray}
\mathcal{G}_L = \frac{2\kappa^2}{V}\int dV \left[\left(\bm{H} -
\bm{H}_0\right)^2 + \kappa^2\left|\bm{\jmath}\right|^2\right].
\label{eq_L1}
\end{eqnarray}
Notice that this expression is a particular case of
Eq.~(\ref{GLem}) which is obtained by putting $|\Psi|^2 = 1$
everywhere inside the disk. In the presence of $L$ vortices,
situated at $\bm{\rho}_i$ $\{i = 1,\, 2,\, ... ,\, L\}$, the
London equation can be written as
\begin{eqnarray}
\bm{J}= \frac{d}{\kappa^2}\big( \bm{\nu} - \bm{A} \big),
\label{eq_L2}
\end{eqnarray}
where
\begin{eqnarray}
\bm{\nu} = \sum_{i=1}^{L}\big[\Phi(|\bm{\rho}-\bm{\rho}_i|)-
\Phi(|\bm{\rho}-(R/\rho_i)^2 \bm{\rho}_i|) \big], \label{eq_L4}
\end{eqnarray}
with $\bm{\rho}_i = (x_i,\,y_i)$ the position of the vortices,
$\bm{J} = \int_0^d dz \bm{\jmath} \approx \bm{\jmath} \, d$, and
$\bm{\Phi}(\rho) = \hat \phi / \rho $. The vortex images at
$(R/\rho_i)^2 \bm{\rho}_i$ appear in Eq.~(\ref{eq_L4}) in order to
fulfill the boundary condition~\cite{Buzdin} $\bm{J}(R) \cdot
\hat\rho = 0$. Instead of writing Eq.~(\ref{eq_L2}) for the vector
$\bm{J}$, one may use the streamline function, $g(\bm{\rho})$,
related to the supercurrent by $\bm{J} = \nabla \times (\hat z g)$
($g(\bm{\rho})$ can be regarded as a local magnetization in the
thin film.~\cite{pehb168}) At the boundary $g(R,\phi)= {\rm
const}$, but, as the value of this constant is arbitrary, one can
impose $g(R,\phi)=0$. Therefore, Eqs.~(\ref{eq_L2})
and~(\ref{eq_L4}) can be expressed as,
\begin{eqnarray}
 g(\bm{\rho})  =  \frac{d}{\kappa^2} & \Big[ & \sum_{j=1}^L
\ln\left(\frac{|\bm{\rho} - (R/\rho_j)^2 \bm{\rho}_j|}
 {|\bm{\rho} - \bm{\rho}_j|}
\frac{\rho_j}{R}\right)  \nonumber \\
& & -  \frac{H_0}{4}\left(R^2- \rho^2\right)\Big]. \label{eq_L7}
\end{eqnarray}
Notice that Eq.~(\ref{eq_L2}) can also be understood as the
limiting case of the GL equations if one considers $|\Psi| = 1$
and $\nabla \theta = \bm{\nu}$. Therefore, while vortices are well
apart from each other (and also the boundary), there exists a
relation between the streamline function defined above and the
phase of the order parameter in the GL theory, i.e., one can
define a complex function of which the real and imaginary parts
are proportional to $g(\bm{\rho})$ and $\theta$.~\cite{PRB22_1200}

Since in our case ($\Lambda = \lambda^2 / d \gg \xi \gg d$),
demagnetization effects can be neglected\cite{Baelus_PRB03} and
one may write Eq.~(\ref{eq_L1}) as
\begin{eqnarray}
\mathcal{G}_L & = & 
\frac{2\kappa^4}{V d} \int d^2\rho \,\left|\bm{J} \right|^2 =
\frac{2\kappa^4}{V d}\int d^2\rho \,g(\bm{\rho})
\hat z \cdot \nabla\times\bm{J} \nonumber\\
& = & \frac{2\kappa^2}{V}\left[2\pi \sum_{i=1}^L g(\bm{\rho}_i)-
H_0\int d^2\rho \,g(\bm{\rho})\right],
 \label{eq_L3}
\end{eqnarray}
where  the integration is performed along the thin film plane, $z
= 0$. Substituting Eq.~(\ref{eq_L7}) in this formula, and after
some algebraic manipulation, the London energy is expressed by
\begin{eqnarray}
\mathcal{G}_L & = &
\left(\frac{2}{R}\right)^2\sum_{i=1}^L\sum_{j=1}^L
\ln\left(\frac{r_j |\bm{r}_i- \bm{r}_j/r_j^2|} {|\bm{r}_i -
\bm{r}_j|}
\right)  \nonumber \\
& & - 2H_0\sum_{i=1}^L \left(1- r_i^2\right)+ R^2 H_0^2,
\label{eq_L8}
\end{eqnarray}
where we used $\bm{r}_i = \bm{\rho}_i / R$ to simplify the
notation.

The divergence in Eq.~(\ref{eq_L8}) can be removed by considering
a cut-off, in which for $i = j \to |\bm{\rho}_i - \bm{\rho}_j| = a
\xi $ (in not normalized units) and $a$ is a constant. The final
expression for the London energy can be written as
\begin{subequations}
\label{eqenergy}
\begin{eqnarray}
\mathcal{G}_L \! = \! \sum_{i=1}^L\left( \epsilon_i^{\rm self} +
\epsilon_i^{\rm shield} + \sum_{j=1}^{i-1} \epsilon_{ij} \right) +
\epsilon^{\rm core} \!+\! \epsilon^{\rm field}, \label{eq_L10}
\end{eqnarray}
where
\begin{eqnarray}
\epsilon_i^{\rm self} = \left(\frac{2}{R}\right)^2
\ln\left(1-r_i^2\right) \label{eq_L10a}
\end{eqnarray}
is the interaction energy between the $i^{\rm th}$ vortex and the
radial boundary of the superconductor,
\begin{eqnarray}
\epsilon_i^{\rm shield} = - 2H_0\left(1- r_i^2\right)
\label{eq_L10b}
\end{eqnarray}
represents the interaction between the $i^{\rm th}$ vortex and the
shielding currents, and
\begin{eqnarray}
\epsilon_{ij} = \left(\frac{2}{R}\right)^2 \ln\left[\frac{(r_i
r_j)^2- 2\bm{r}_i\cdot\bm{r}_j+1} {r_i^2- 2\bm{r}_i\cdot\bm{r}_j
+r_j^2}\right] \label{eq_L10c}
\end{eqnarray}
\end{subequations}
is the repulsive energy between vortices $i$ and $j$. Finally,
$\epsilon^{\rm core} = (2/R)^2N\ln (R/a)$ and $\epsilon^{\rm
field} = R^2 H_0^2$ are the energies associated with the vortex
cores and the external magnetic field, respectively.

Notice that $\mathcal{G}_L$ allows one to treat the vortices as
particles. Therefore, simulation techniques appropriate for
systems of particles may be performed in order to find, for
example, the ground state of the
system.~\cite{PRB49b,prb51_7700,pre67_021608} In this sense, the
vortex system behaves (in the London approximation) similar to a
two dimensional system composed of equally charged particles
interacting through a repulsive logarithmic potential placed in
parabolic potential well.~\cite{JPCM9_5383,pre60_4743}
Nevertheless, there is a fundamental difference between these two
systems: the vortex system is confined to the disk of radius $R$
and the influence of the surface on the energy is clear from the
terms containing vortex images, i.e., $\epsilon_i^{\rm self}$ and
$\epsilon_{ij}$. Notice also that $\epsilon^{\rm core}$ arises
from the cut-off procedure and is therefore strongly dependent on
the cut-off value $a\xi$ (we adopted $a = 1$ in the results shown
below). The actual energy associated with vortex cores and with
the spatial variation of the superconducting electron density
($\left|\psi(\bm{\rho})\right|^2$) should be evaluated by using
the GL theory.

A thin disk with $L$ vortices was simulated by using
Eq.~(\ref{eq_L10}). To investigate (meta-)stable states close to
the equilibrium, we employed a procedure similar to the one
described in Ref.~\onlinecite{prb51_7700}. First $L^{\prime}$
vortices were distributed randomly inside the disk. Then, a Monte
Carlo (MC) technique was used to make the system wander in the
configurational space and arrive at a neighborhood of some minimum
of $\mathcal{G}_L$. After typically $10^4$ MC steps, we perform a
molecular dynamics (MD) simulation starting from the final MC
configuration. The final (meta)-stable state is achieved after
about $10^6$ MD steps. In order to find the ground state (or
states with energies very close to it) this trial procedure was
repeated more than $1000$ times, each time starting with a
different random distribution of $L'$ vortices at a given magnetic
field $H_0$.

To implement the MD we time integrated the Bardeen-Stephen
equation of motion~\cite{BardeenStephen}
\begin{eqnarray}
\eta\frac{d \bm{\rho}_i}{dt} = \bm{F}_i, \label{eq_L11}
\end{eqnarray}
where $i$ is the label of the $i^{\rm th}$ vortex, $\eta$ is the
viscous drag coefficient $\eta \sim \Phi_0 H_{c2} / \rho_n c^2$
(where $\rho_n$ is the normal state resistivity). The forces
acting on each vortex were obtained from $-\nabla_k
\mathcal{G}_L(\bm{\rho}_i,\bm{\rho}_j)$, where $\mathcal{G}_L$ is
given by Eq.~(\ref{eq_L10}) and $-\nabla_k$ is the gradient with
respect to the coordinate $\bm{\rho}_k$. This yields a force per
unit of volume,
\begin{subequations}
\label{eqforce}
\begin{eqnarray}
\bm{F}_i = \bm{F}_i^{s} + \sum_{  k = 1  \atop{k \ne i} }^L
\bm{F}_{i,k}^{int}, \label{eq_L11a}
\end{eqnarray}
which we express in units of $H_c^2 / 8\pi\xi$ . Above, the first
term describes the vortex interaction with the current induced by
the external field and with the interface,
\begin{eqnarray}
\bm{F}_i^{s} = \left(\frac{2}{R}\right)^3\left(\frac{1}{1 -
r_i^2}-\frac{H_0 R^2}{2}\right)\bm{r}_i, \label{eq_L11a1}
\end{eqnarray}
and the second, the vortex-vortex interaction,
\begin{eqnarray}
\bm{F}_{i,k}^{int} =
\left(\frac{2}{R}\right)^3\left(\frac{\bm{r}_i-\bm{r}_k}{\left|\bm{r}_i
- \bm{r}_k\right|^2}- r_k^2 \frac{r_k^2\bm{r}_i - \bm{r}_k}{\left|
r_k^2\bm{r}_i - \bm{r}_k \right|^2}\right). \label{eq_L11a2}
\end{eqnarray}
\end{subequations}
The simple Euler method was used to accomplish the time
integration, but adopting a $\delta t$ value small enough to avoid
large variations of the vortex positions between two consecutive
steps. Moreover, the dynamical matrix (the Hessian matrix of
$\mathcal{G}_L$), whose elements are given by
\begin{eqnarray}
\frac{\partial^2 \mathcal{G}_L}{\partial \rho_{\alpha,i}\partial
\rho_{\beta,j}}, \label{eq_L11b}
\end{eqnarray}
was calculated for the final vortex configuration. In this
equation, the Greek indexes stand for the components of the vector
$\bm{\rho_i}$, while the Italic indexes are the labels for the
vortices. The computation of the dynamical matrix eigenvalues
allowed us to tell whether the given state was stable or unstable
(for a stable state all the dynamical matrix eigenvalues must be
non-negative). Unstable states were discarded.

One difficulty in simulating this system is the fact that both
$\mathcal{G}_L$ and the forces acting on the vortices diverge at
the disk edge. To overcome this, during the MD simulation whenever
a vortex was at a distance less than $\xi$ from the disk edge, it
is taken out from the system, i.e., this vortex disappears.
Therefore, the final number of vortices may not be the same as in
the beginning. This does not lead to any serious concern, since we
collect all the final results from each trial and sort them in
ascending order of energy. It also allows to compare energies of
systems containing different number of vortices for the same
external magnetic field and investigate which of them correspond
to the lower energy, i.e., is the ground state.

\section{Low $L-$states: Vortex Molecules}
\label{seciii}

\begin{figure}[!tbp]
\centering
\subfigure[]
{ \label{Fig_2a}
\includegraphics[width=0.2\textwidth]{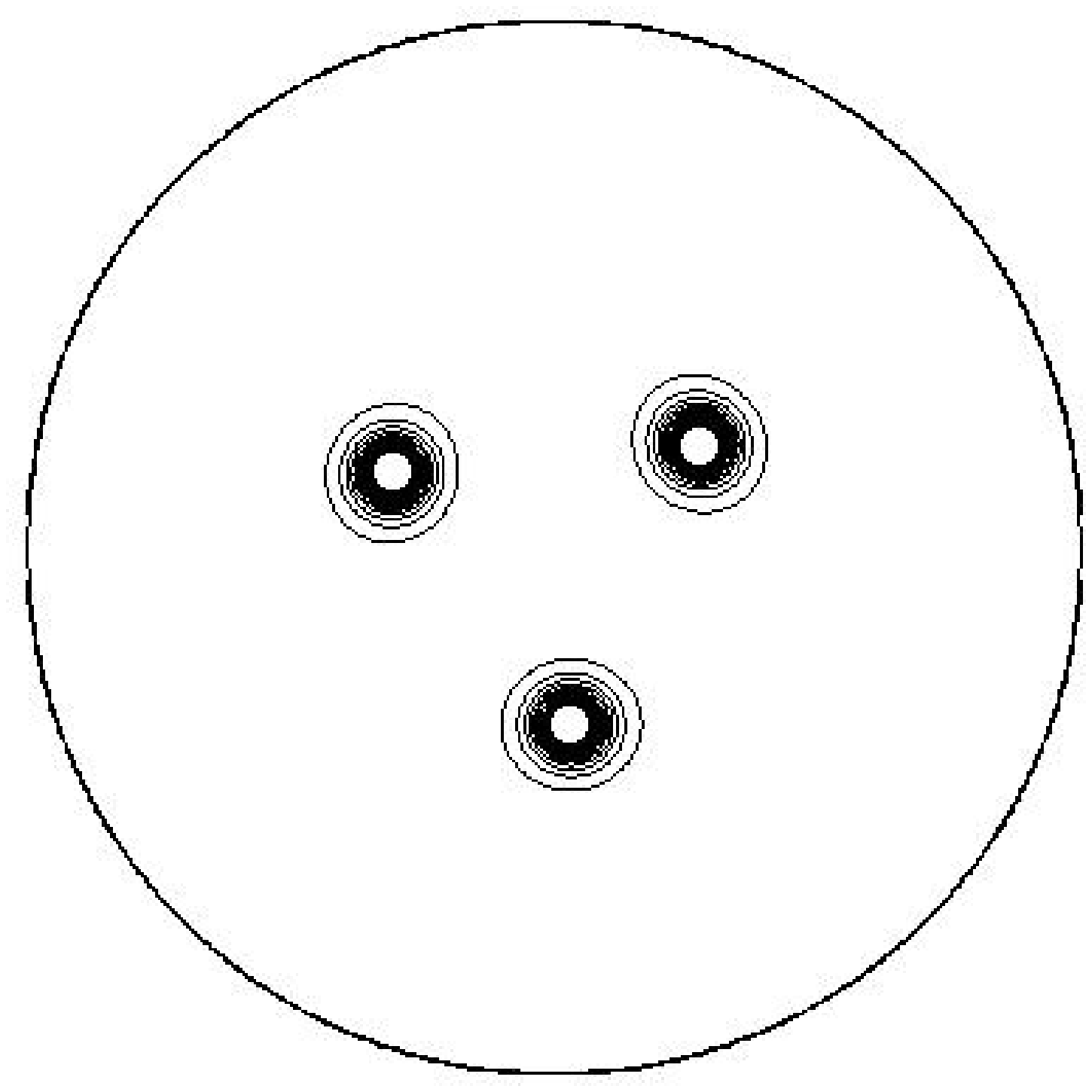}
} 
\subfigure[]
{ \label{Fig_2b}
\includegraphics[width=0.2\textwidth]{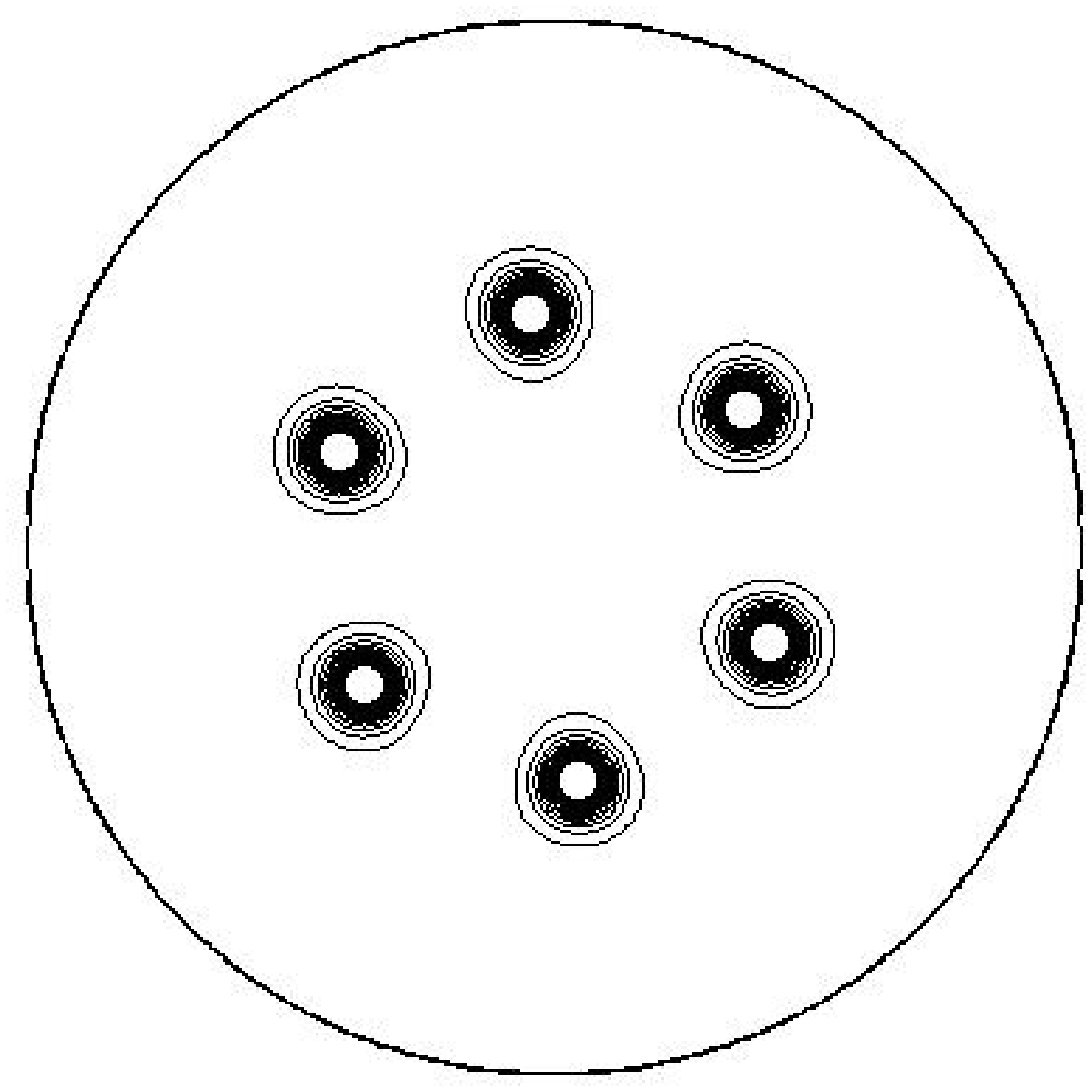}
} \vspace{0.5cm} \subfigure[] { \label{Fig_2c}
\includegraphics[width=0.2\textwidth]{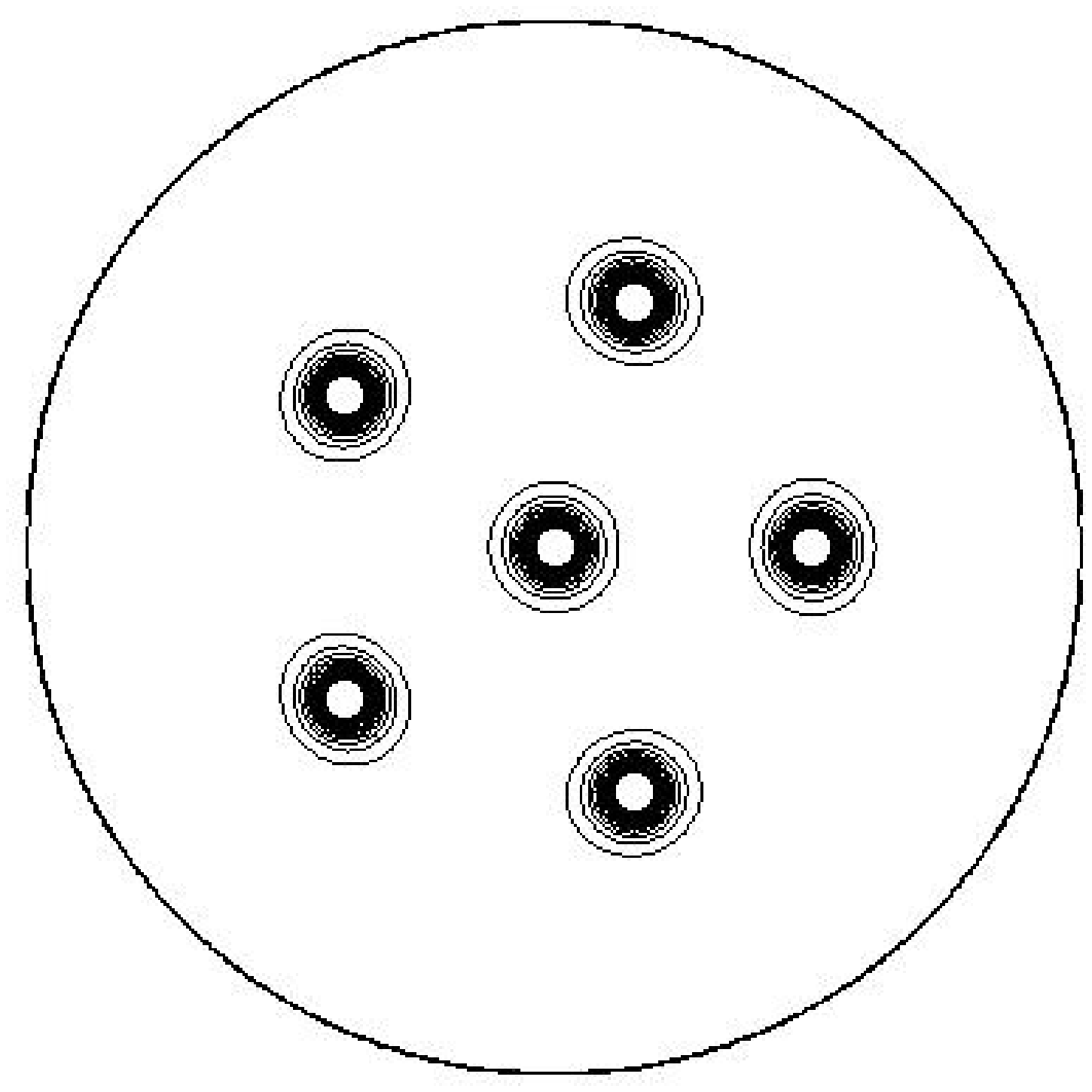}
} 
\subfigure[] { \label{Fig_2d}
\includegraphics[width=0.2\textwidth]{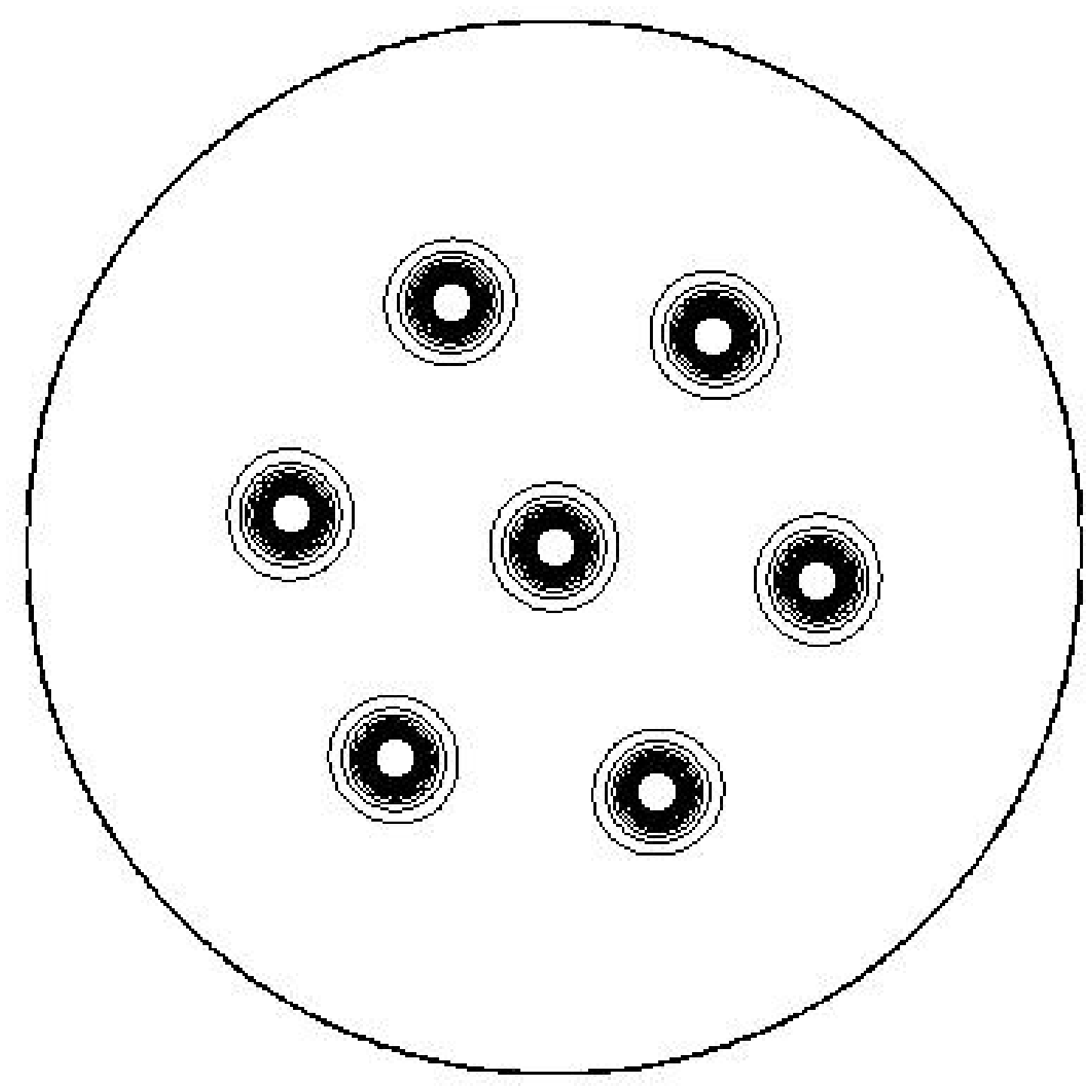}
} \subfigure[]{\label{Fig_2e}
\includegraphics[width=0.45\textwidth]{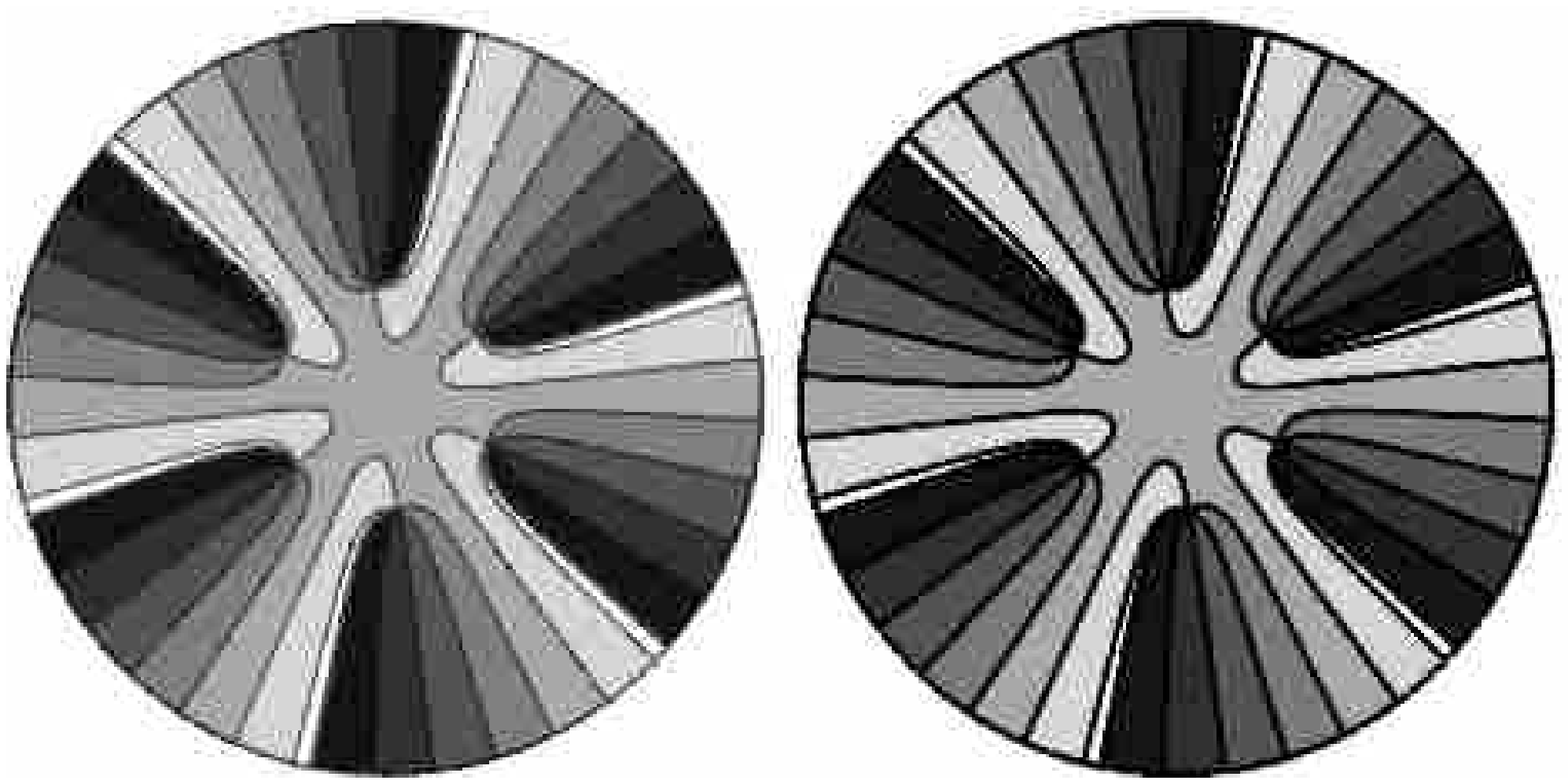}}
 \vspace{0.0cm}
\caption{Vortex configurations for $L = (3)$ and $H_0 =
0.007$~(a), $L =(6)$ and $H_0 = 0.01$~(b), $L =(1,5)$ and $H_0 =
0.01$~(c), and $L = (1,6)$ and $H_0 = 0.011$~(d). The black lines
are the contour lines of $|\Psi(\bm{r})|^2$, whereas the white
circles indicate the position of the vortices according to the
London approximation. In (e) we show the phase of the order
parameter for the $L = (6)$ state at $H_0 = 0.022$ obtained from
the GL equations (on the left) and from the London approximation
(on the right). \label{Fig_2} }
\end{figure}

In this section we present the results calculated from the GL and
London theories for low $L-$states for a thin disk of radius $R =
50\xi$. A comparison between ground states in the GL theory and
the London approximation was done in
Ref.~\onlinecite{Baelus_PRB03}, for the case of a small disk
radius (i.e. $R = 6\xi$). In that case, it was not possible to
study multivortex configurations for $L-$states above $L = 14$
since the calculated GL results showed only giant vortices.
Moreover, above $L = 26$ the disk was driven to the {\it normal
state}. In the present case, multivortex configurations are
obtained for much higher $L-$states. This enabled us to compare
large multivortex configurations calculated by both the GL theory
and the London approximation, and investigate the transition to
the Abrikosov lattice.

For $L = 1$ to $L = 9$, the lowest energy configurations consist
of vortices distributed in regular polygons with $0$ or $1$ vortex
in the center of the disk. This means that not many meta-stable
states are close to the ground state, which makes the job of
finding low energy configurations easier. In the London limit,
this reduces Eqs.~(\ref{eq_L10})--(\ref{eq_L10c}) to a simple
form, which depends on only one free
parameter,~\cite{Buzdin,Baelus_PRB03} i.e., the radius of the ring
which circumscribes the polygon, $\rho_{\rm ring}$. The
minimization problem is then straightforward. We also obtained the
positions of the vortex ring by finding the roots of
\begin{eqnarray}
\!\!\frac{1}{1-r^2} - h + \frac{N \pm 1}{2r^2} 
 - \! \sum_{n = 1}^{N-1}\! \frac{r^2 - \cos \phi_n}{1 + r^4 - 2r^2\cos
\phi_n} = 0 , \label{eqA1}
\end{eqnarray}
which follows from the balance of forces acting on each vortex
[cf. Eq~(\ref{eqforce})].~\cite{Buzdin} Here $N$ is the number of
vortices on the ring (or the number of sides of the polygon), $r =
\rho_{\rm ring} / R$, $\phi_n = 2\pi n/N$, $h = H_0R^2/2$ and the
plus (minus) sign should be taken if there is one (zero) vortex in
the center of the disk.

A comparison between the calculated GL and London vortex
configurations is depicted in Fig.~\ref{Fig_2}. The states $L = 3$
(Fig.~\ref{Fig_2a}), $L = (6)$ (Fig.~\ref{Fig_2b}), $L = (1,5)$
(Fig.~\ref{Fig_2c}), and $L = 7$ (Fig.~\ref{Fig_2d}) were obtained
at $H_0 = 0.007$, $H_0=0.01$, $H_0=0.01$, and $H_0 = 0.011$,
respectively. The vortex positions practically coincide for the
same configurations in both theories.

The agreement between the vortex positions yielded by both
theories (at $H_0 \ll H_{c2}$) is related to the fact that the
phase of the order parameter, $\theta$, is well described as the
imaginary part of the complex function
\begin{eqnarray}
\Omega  =  \sum_{j=1}^L \ln\left[\left(\frac{\zeta - (R/\zeta_j)^2
\zeta_j} {\zeta - \zeta_j}\right) \frac{\rho_j}{R}\right] -
\frac{H_0}{4}\left(R^2- \rho^2\right)\!\!, \,\,\label{eq_13}
\end{eqnarray}
for sufficiently small magnetic fields,~\cite{PRB22_1200} where
$\zeta = \rho e^{i\phi} = x + i y$ is the representation of the
vector $\bm{\rho}$ in the complex $\zeta-$plane. But $(d/\kappa^2)
Re\{\Omega\}$ is simply the streamline function [cf.
Eq.~(\ref{eq_L7})] calculated in the London limit. That is greatly
responsible for the fact that $\rho_{\rm ring}$ is virtually the
same in both theories for $H_0 \ll H_{c2}$. Fig.~\ref{Fig_2e}
presents the numerically calculate phase of the order parameter
(left) and the theoretical one obtained from the imaginary part of
Eq.~(\ref{eq_13}) (right) for the state with $L = 6$ at $H_0 =
0.022$.

\begin{figure}[!tbp]
\centering
\includegraphics[width=0.495\textwidth]{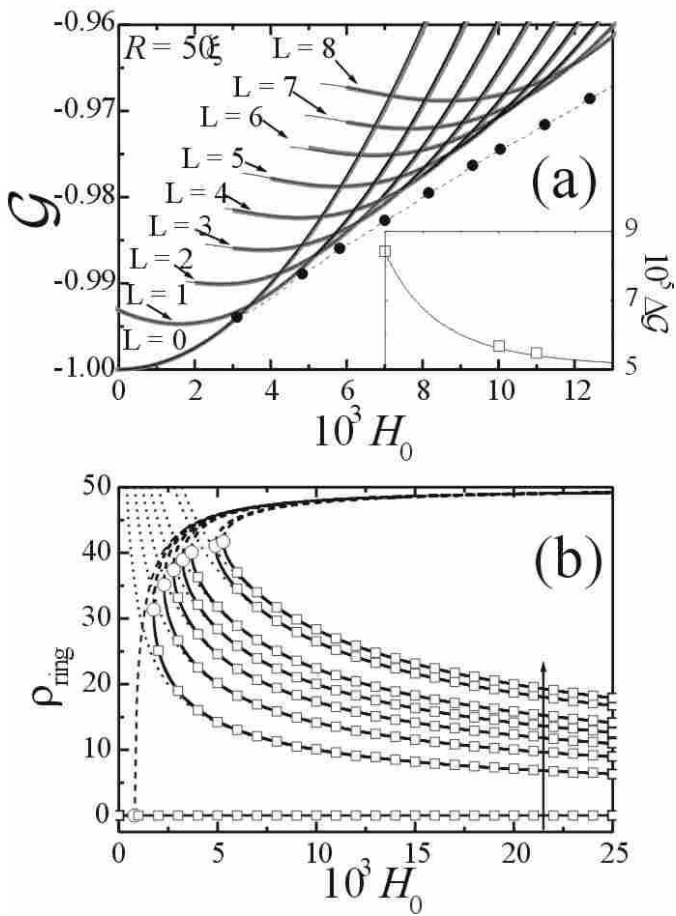}
 \vspace{0.0cm}
\caption{(a): The GL (thick lines) and the improved London (thin
lines) free energies as function of the applied field $H_0$ for
low $L-$states. The $L = (1,5)$ state has slightly lower energy
than the $L = (6)$ state, as seen in the inset, where the lines
and the squares show the difference between the $L = (6)$ and $L =
(1,5)$ energies in the London limit and in the GL theory,
respectively. The usual London energy (where we added $-1$) is
also depicted (dashed lines) for comparison. The solid circles
show the points at which the usual London energy predicts a
transition from $L$ to $L + 1$. (b): the GL (open squares) and
London (solid lines) radial position of the vortices in the ring
($\rho_{\rm ring}$) as function of the magnetic field for the $L =
1$, $(2)$, $(3)$, $(4)$, $(5)$, $(6)$, $(1,6)$, $(1,7)$ states.
The arrow indicates the direction of increasing $L$. For each
configuration (in the London limit) the vortex position at which
the vortex ring is unstable (dashed lines) and the onset field
from which stability occurs (open circles) are depicted. The
radial positions of the vortex ring when the boundary induced
`vortex images' are neglected are shown by the dotted lines for
comparison. \label{Fig_1} }
\end{figure}

The dependence of $\rho_{\rm ring}$ upon $H_0$ is shown in
Fig.~\ref{Fig_1}(b) obtained within the GL (squares) and the
London limit (solid line) for the $L = 1$, $(2)$, $(3)$, $(4)$,
$(5)$, $(6)$, $(1,6)$, $(1,7)$ states. Both theories predict the
same values of $\rho_{\rm ring}$ and, thus, the same stable
configurations, as function of $H_0$. Fig.~\ref{Fig_1}(b) also
shows the radial position over which a given regular polygon
configuration is not stable (dashed lines) as function of $H_0$
(obtained in the London limit). The magnetic field in which the
stable and unstable $\rho_{\rm ring}$ lines start to departure
from each other (open circles) mark the onset of stability for
each configuration. The unstable $\rho_{\rm ring}$ lines merge to
\begin{eqnarray}
R\sqrt{1 - \frac{2}{H_0 R^2}}. \label{eq_L14}
\end{eqnarray}
This is simply the position after which the attractive force
acting on each vortex by its own image becomes larger than the
force produced by the shielding currents (which pulls the vortices
inside), as can be easily demonstrated from Eqs.~(\ref{eq_L11a})
and~(\ref{eq_L11a1}) for one vortex. It is also important to take
into account the vortex interaction with the disk edge for
sufficiently low fields. This can be noticed from the difference
between the stable $\rho_{\rm ring}$ and the dotted lines in
Fig.~\ref{Fig_1}(b), which depicts the position at which the
respective regular polygon configuration would sit if there were
no vortex images [from Eq.~(\ref{eqA1}) in the absence of vortex
images, $\rho_{\rm ring}$ would be given by $\sqrt{(N \pm
1)/H_0}$, where the $+$ ($-$) sign should be considered for one
(zero) vortex in the center].

The free energies within the GL (thick lines) theory and the usual
London limit (dashed line) are depicted in Fig.~\ref{Fig_1}(a) for
$L = 0 \to 8$ as a function of the applied magnetic field $H_0$.
The energy calculated within the London limit (with $a = 1$)
starts to departure from the GL results as soon as $L = 1$. This
is mainly due to the fact that the usual London theory neglects
the spatial variation of $|\Psi^2|$. When the magnetic field
increases, the ground state changes by the addition of one vortex,
i.e, $L = 0 \to 1 \to 2 ... \to 8$ (for the London limit these
transitions are marked by the filled circles). For disks with
small radius the GL theory predicts that $L = 2 \to 6$ states do
not have a vortex in the center of the
disk.~\cite{PRL81v,Baelus_PRB03} Such a central vortex appears in
the $L = 7 \to 9-$states.  In contrast, for the present large disk
case ($R = 50\xi$), the GL theory and the London approximation
yield five vortices arranged in a regular pentagon with one in the
center of the disk for $L = 6$. The state with six vortices in a
regular hexagon has a slightly higher energy [the difference in
energy is depicted in the inset of Fig.~\ref{Fig_1}(a)].

In an effort to remedy the differences in the energy between the
GL and the usual London results we considered the contribution of
the vortex cores energies to the London energy. As long as
vortices are well separated and $H_0 \ll 1$ ($|\Psi|^2 \approx 1$
far from the vortex cores), Eq.~(\ref{GLem}) can be approximately
given by the London energy. In this limit the depreciation of
$|\Psi|^2$ around the vortex cores can be approximated by the
superposition of some function which varies from $0$ to $1$ within
$|\bm{\rho} - \bm{\rho}_i| \sim \xi$. Such extensions of the
London theory were previously
considered~\cite{Clem75a,Brandtreview} for infinite
superconducting systems, e.g., by using $\left|\Psi\right|^2 =
\left|\bm{\rho} - \bm{\rho}_i \right|^2 / \left( \left|\bm{\rho} -
\bm{\rho}_i \right|^2 + 2\xi^2\right)$ close to the core of the
vortex at $\bm{\rho}_i$. We used this expression into
Eq.~(\ref{GLcore}) in the limit that vortices are far apart, i.e.,
for low $L$ values, where we can make use of the superposition
principle. First,  Eq.~(\ref{GLcore}) can be written as
\begin{eqnarray}
 \mathcal{G}_{\rm core}
  =  -1 \!+\! \frac{1}{\pi R^2}\int\left[ \left(1 -
 \left| \Psi \right| ^{2} \right)^2
 \!+ \!2\left(\nabla \left| \Psi
\right|\right)^{2}\right] d^2\rho. \label{GLcore2}
\end{eqnarray}
Close to the cores, $1-|\Psi|^2 = 2/\left( \left|\bm{\rho} -
\bm{\rho}_i \right|^2 + 2\right)$ and $\nabla|\Psi| = 2/\left(
\left|\bm{\rho} - \bm{\rho}_i \right|^2 + 2\right)^{3/2}$
(remembering that $\xi = 1$ in our units). Since these expressions
rapidly approach zero, we approximated the integration over the
disk area in Eq~(\ref{GLcore2}) by the sum of integrations around
of the vortex cores. This yields
\begin{eqnarray}
 \mathcal{G}_{\rm core}
& \approx & -1 + L \frac{3}{R^2}. \label{GLcore3}
\end{eqnarray}
We added the above value of $\mathcal{G}_{\rm core}$ to the London
energy, $\mathcal{G}_{L}$, assuming that the vortex core have a
radius $\sqrt{2}\xi$, which yields $a = \sqrt{2}$ in
$\epsilon^{\rm core}$. The resulting improved London energies are
presented in Fig.~\ref{Fig_1}(a) by thin lines for the $L = 1$,
$(2)$, $(3)$, $(4)$, $(5)$, $(1,5)$, $(1,6)$, and $(1,7)$ states.
The agreement between this improved London theory with the GL
results is very good. Such extension of the London limit yields
the region over which each configuration is the ground state with
much more confidence than the usual London limit.

In the above approximation for $\mathcal{G}_{\rm core}$ the
depreciation of the order parameter near the disk edge was
neglected. In order to have an estimate of the behavior of
$|\Psi|^2$ close to $\rho = R$, we may consider the first GL
equation written as
\begin{eqnarray}
-\nabla^2\Psi + \Psi\left(1 - \left|\Psi\right|^2 - \Pi^2\right) =
0,\label{eq1GLm}
\end{eqnarray}
with boundary conditions $\frac{\partial |\Psi|}{\partial
\rho}\big|_{\rho = R} = 0$ and $\hat \rho \cdot \Pi\big|_{\rho =
R} = 0$. Notice that $\Pi = \nabla \theta - \bm{A} = \bm{\nu} -
\bm{A}$ automatically satisfies its boundary condition if
$\bm{\nu}$ is considered within the London limit [cf.
Eq.~(\ref{eq_L4})]. For a giant vortex state, $|\Psi|^2$ is
radially symmetric, and $\bm{\nu} = \hat \phi L / \rho$. For a
regular polygon vortex configuration and after averaging
$\bm{\nu}$ along the angular direction, one finds $\bm{\nu} = \hat
\phi L\Theta(\rho - \rho_{\rm ring})/ \rho$, where $\Theta(x)$ is
the Heaviside step function. Therefore, one may approximate the
superconducting electron density by $|\Psi_{\rm app}|^2 \approx 1
- \left(L / \rho - \rho H_0/ 2 \right)^2$ inside a ring with
internal radius, $R_1$, taken somewhat larger than $\rho_{\rm
ring}$ and external radius smaller than $R - \xi$ (since the term
$\nabla^2\Psi$ in the first GL equation becomes more important
within distances of $\xi$ close to the disk edge). $|\Psi_{\rm
app}|^2$ is minimal at $\rho = R$ and consequently we can use its
value at the boundary in order to estimate when the depreciation
of $|\Psi|^2$ close to the edge becomes important (notice that the
actual $|\Psi|^2$ is higher close to the disk edge than our
approximate result, since there is a correction of order
$\nabla^2\Psi /\Psi$, with $\nabla^2\Psi
> 0$, in this region). We found that a $5\%$ depreciation
in $|\Psi_{\rm app}(R)|^2$ (which would mean $|\Psi(R)|^2 >
0.95$), requires that $H_0 \approx 0.009$ for $L = 0$, $H_0
\approx 0.0098$ for $L = 1$, $H_0 \approx 0.0106$ for $L = 2$,
$H_0 \approx 0.0114$ for $L = 3$, $H_0 \approx 0.0122$ for $L =
4$, $H_0 \approx 0.013$ for $L = 5$, $H_0 \approx 0.0138$ for $L =
6$, $H_0 = 0.0146$ for $L = 7$ and $H_0 = 0.0154$ for $L = 8$,
which are magnetic field values well above the respective regions
where each of theses states are the ground state. Also the order
parameter depreciation close to the disk edge results in a less
rapid increment of the energy of each $L-$state compared with the
energy found within the London limit. But for $H_0 \ll H_{c2}$,
such difference only becomes pronounced at fields well above the
magnetic field region over which the respective $L-$state is the
ground state. Nevertheless, the depreciation of the order
parameter close to the edges is important if one wishes to
understand the entry and exit of vortices in a finite system.

\section{High $L-$states: Abrikosov Lattice}
\label{seciv}

\begin{figure}[!tbp]
\centering
\includegraphics[width=0.5\textwidth]{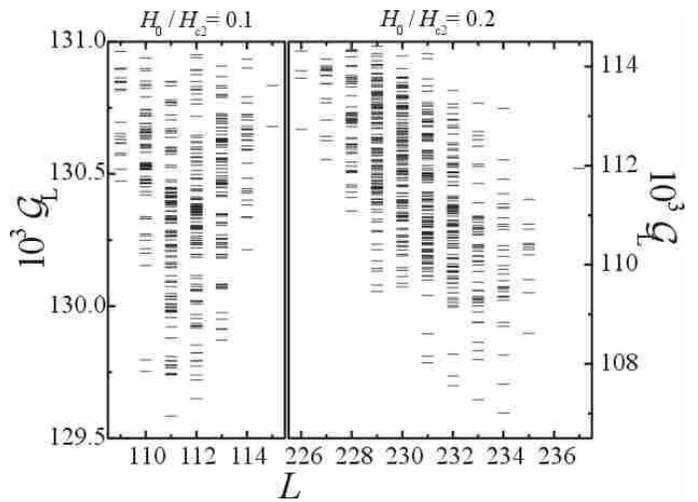}
 \vspace{0.0cm}
\caption{Energies of the meta-stable states ($L = 109 \to 115$ and
$L = 226 \to 237$) obtained from simulations within the London
limit at $H_0 / H_{c2} = 0.1$ (left) and $H_0 / H_{c2}= 0.2$
(right). The energy difference between two different $L-$states is
comparable to the energy difference between distinct
configurations at the same $L-$state. \label{Fig_Eh125} }
\end{figure}

For large values of the vorticity an Abrikosov lattice appears in
the interior of the disk. In this section we will consider $H_0
> 0.03$ and investigate from which value of $L$ the Abrikosov
lattice start to occupy a substantial area in the center of the
disk.

One difficulty which arises when studying the high $L-$states is
due to the fact that the energy difference between two different
$L-$states and the energy difference between distinct
configurations with the same $L$ can be comparable and very small.
This is illustrated in Fig.~\ref{Fig_Eh125}, where the energy of
the meta-stable states obtained in the London limit at $H_0 = 0.1$
and at $H_0 = 0.2$ are shown. For instance, the difference between
the two lowest energy $L = 110$ and $L = 112$ states is less than
$10^{-4}$. At $H_0 = 0.1$ ($H_0 = 0.2$) we found that a vortex
configuration with $L = 111$ ($L = 234$) has the lowest London
energy. Of course it is always possible that configurations with
lower London energies have not been reached by our simulations
(due to the fact that we have a finite number of trials, i.e., we
made typically $1000$ trials). Nevertheless, the small difference
in the energies give us confidence that some of these
configurations are at least very close to the true ground state
within the London limit. Moreover, at such high $L$ values, it is
expected that the energy yielded by the London approximation
differs considerably from the more realistic results obtained from
the GL theory.

\begin{figure}[!tbp]
\centering
\begin{minipage}[t]{0.47\linewidth}
\includegraphics[width=1\textwidth]{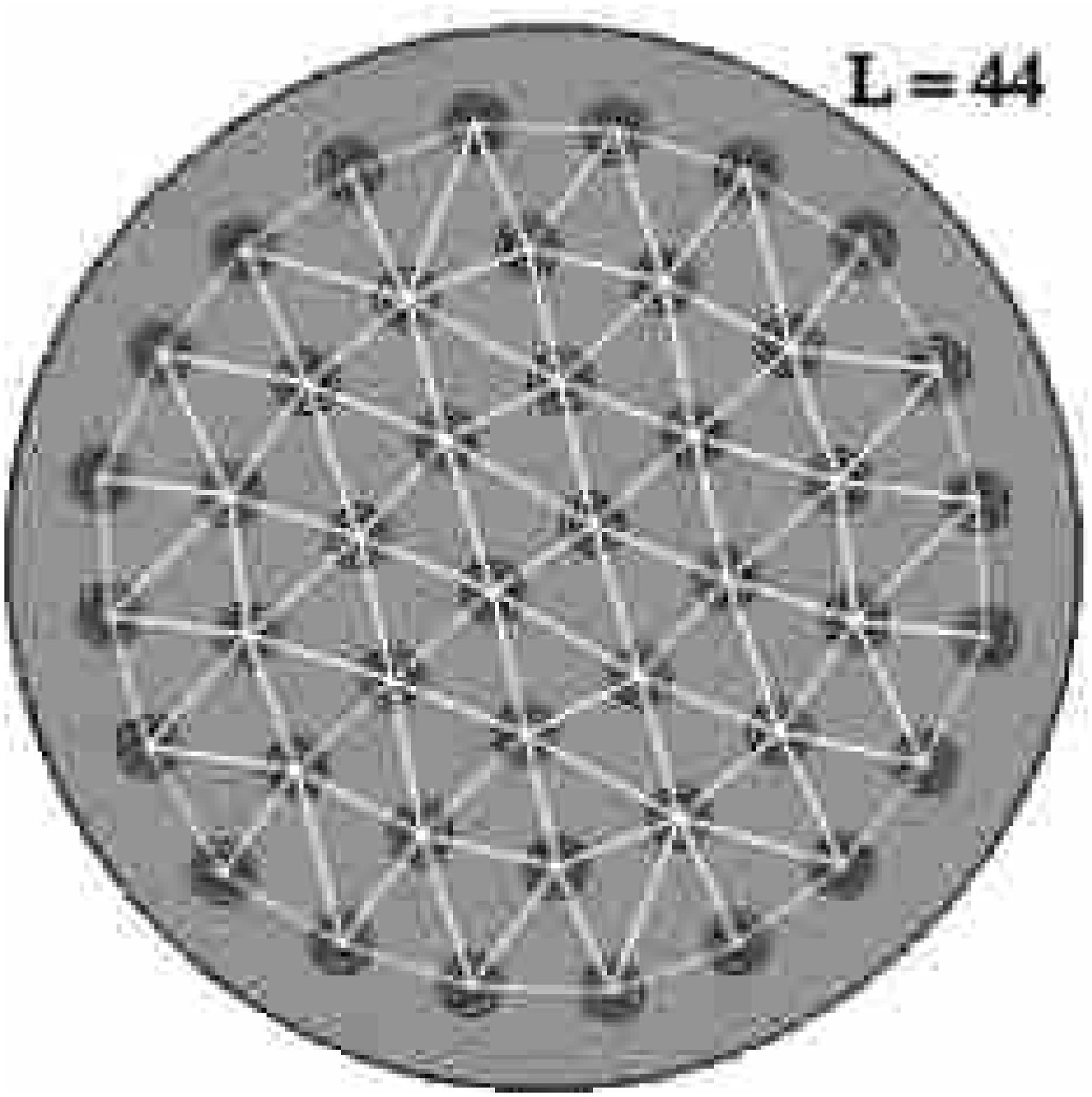}
\end{minipage}
\begin{minipage}[t]{0.47\linewidth}
\includegraphics[width=1\textwidth]{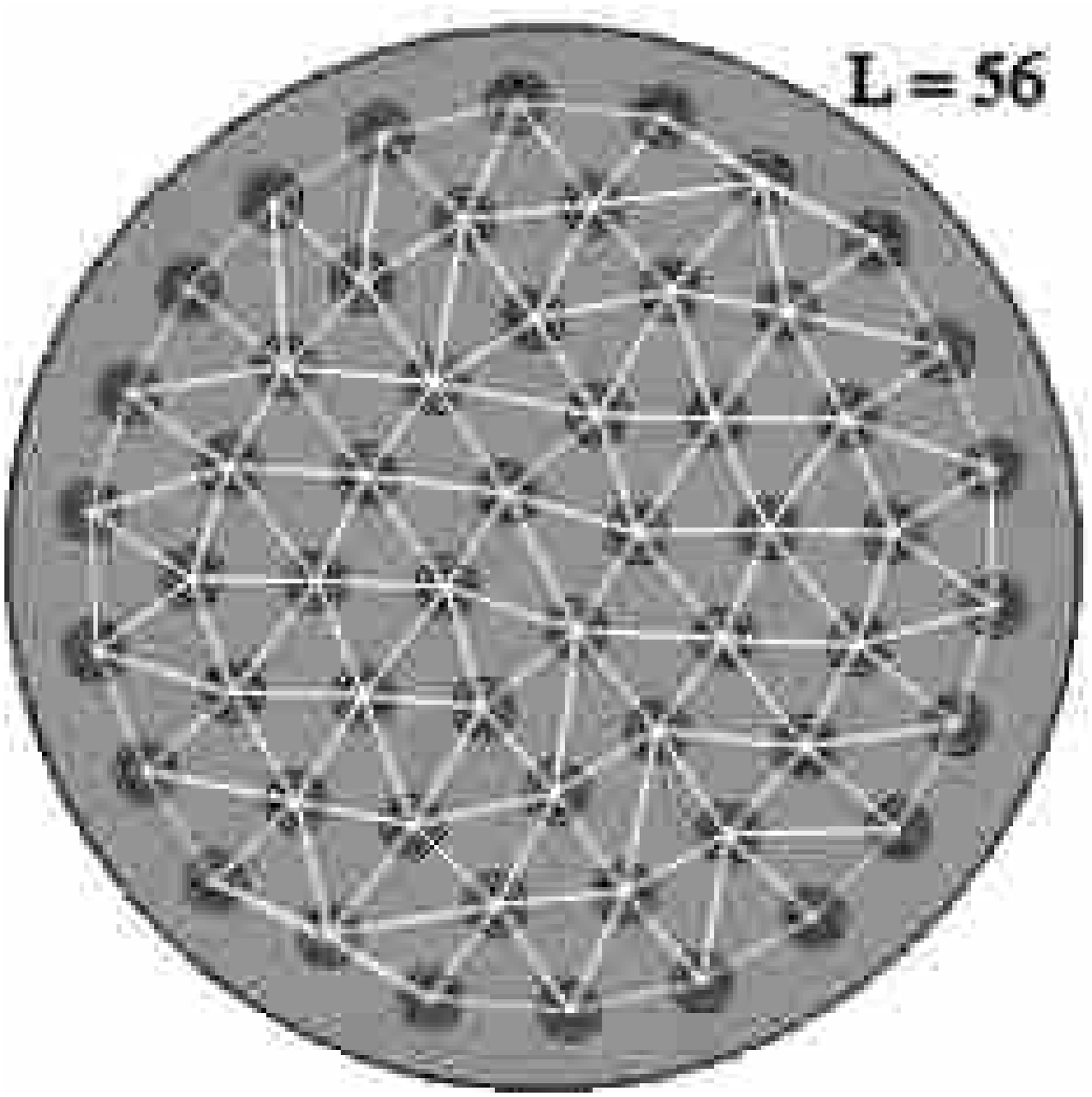}
\end{minipage}
\begin{minipage}[t]{0.47\linewidth}
\includegraphics[width=1\textwidth]{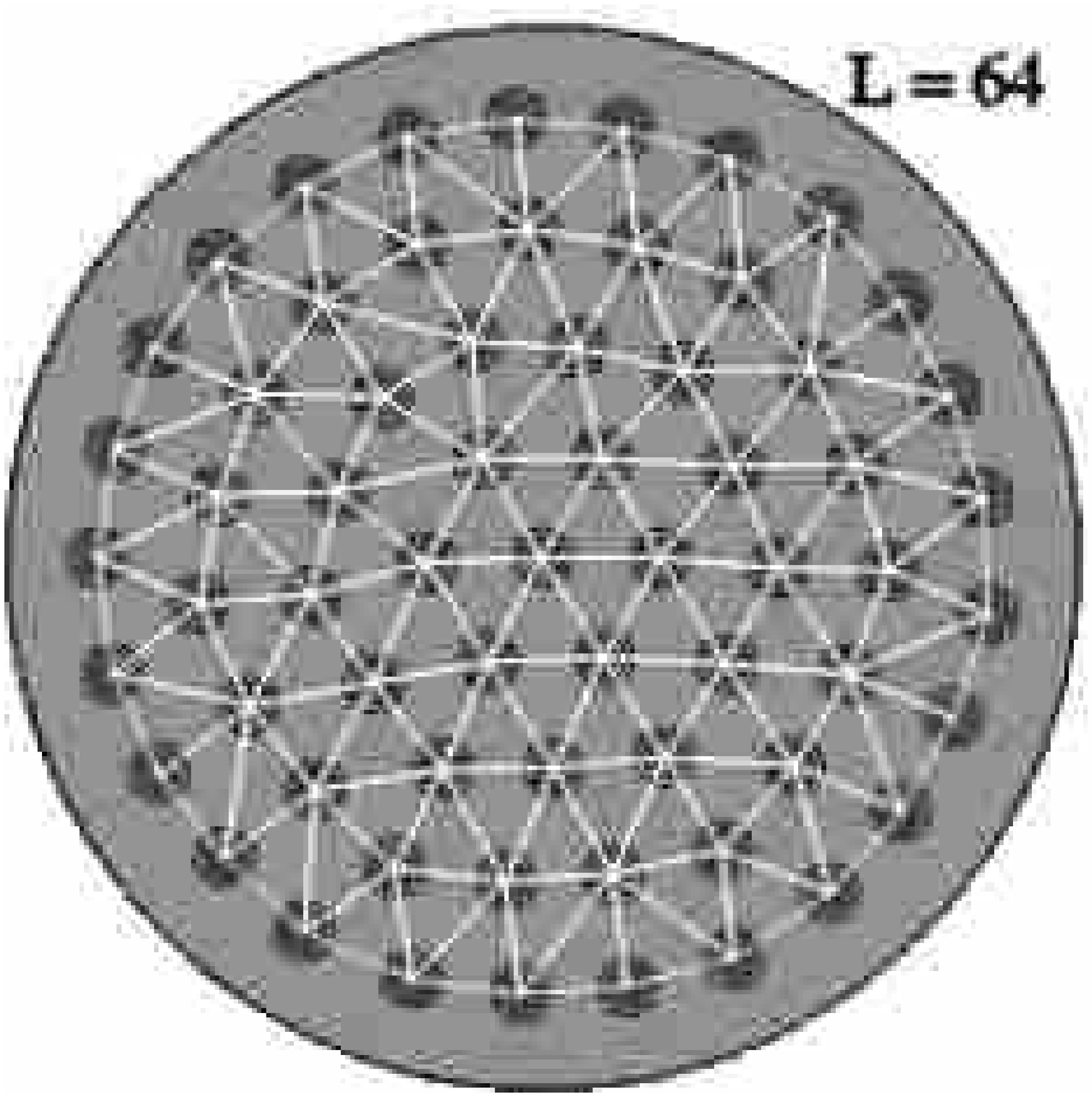}
\end{minipage}
\begin{minipage}[t]{0.47\linewidth}
\includegraphics[width=1\textwidth]{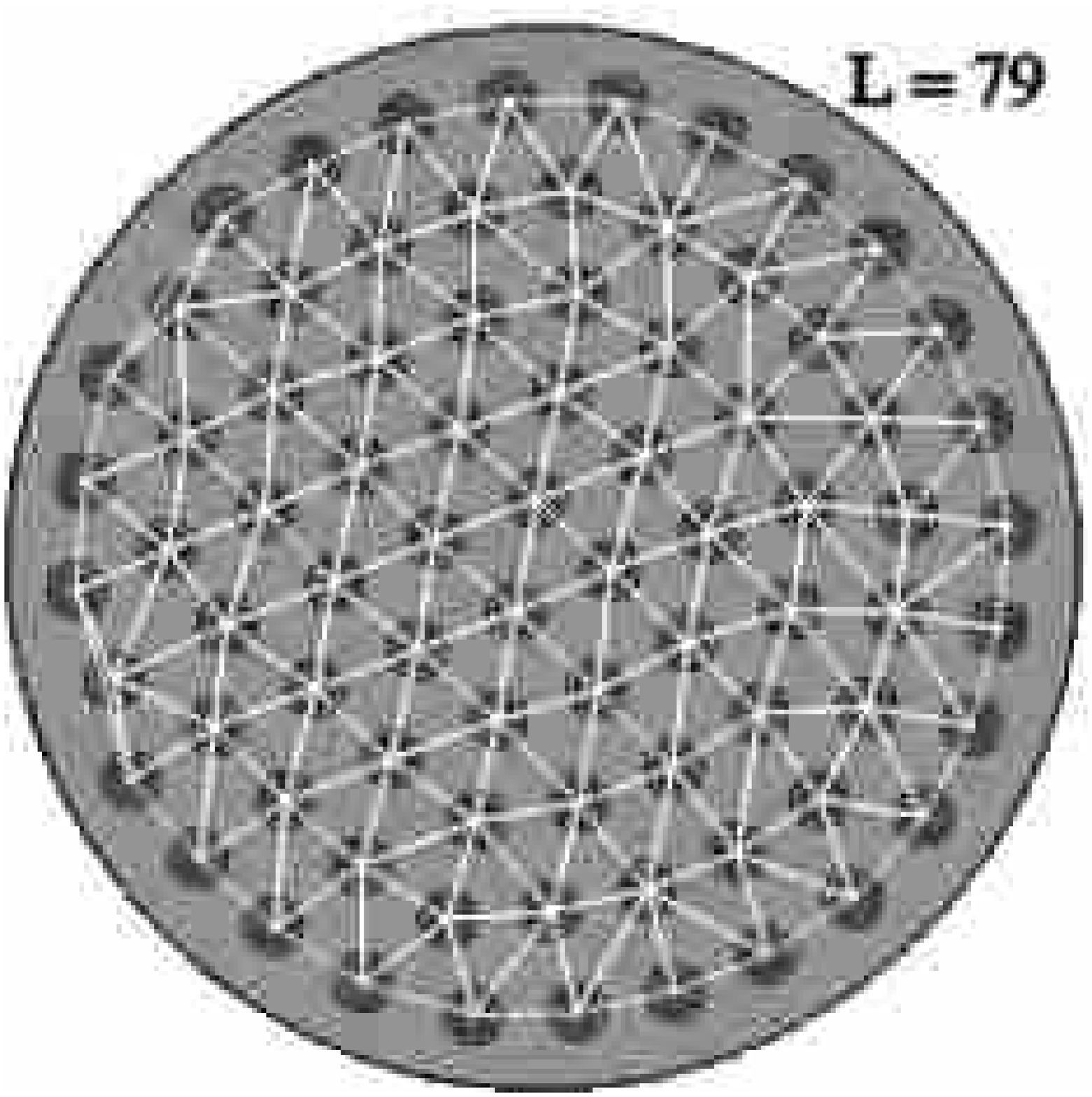}
\end{minipage}
\begin{minipage}[b]{0.47\linewidth}
\includegraphics[width=1\textwidth]{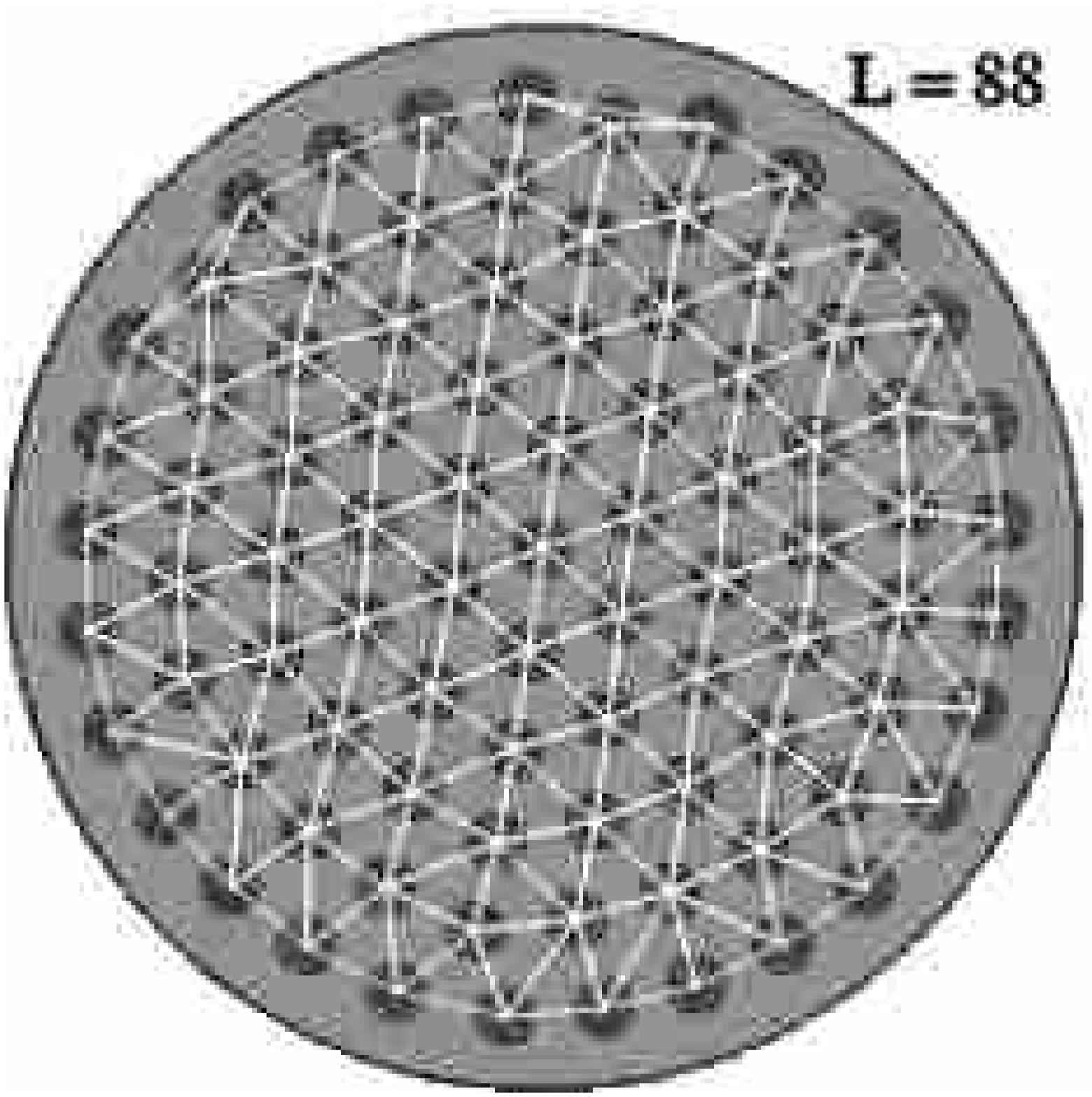}
\end{minipage}
\begin{minipage}[b]{0.47\linewidth}
\includegraphics[width=1\textwidth]{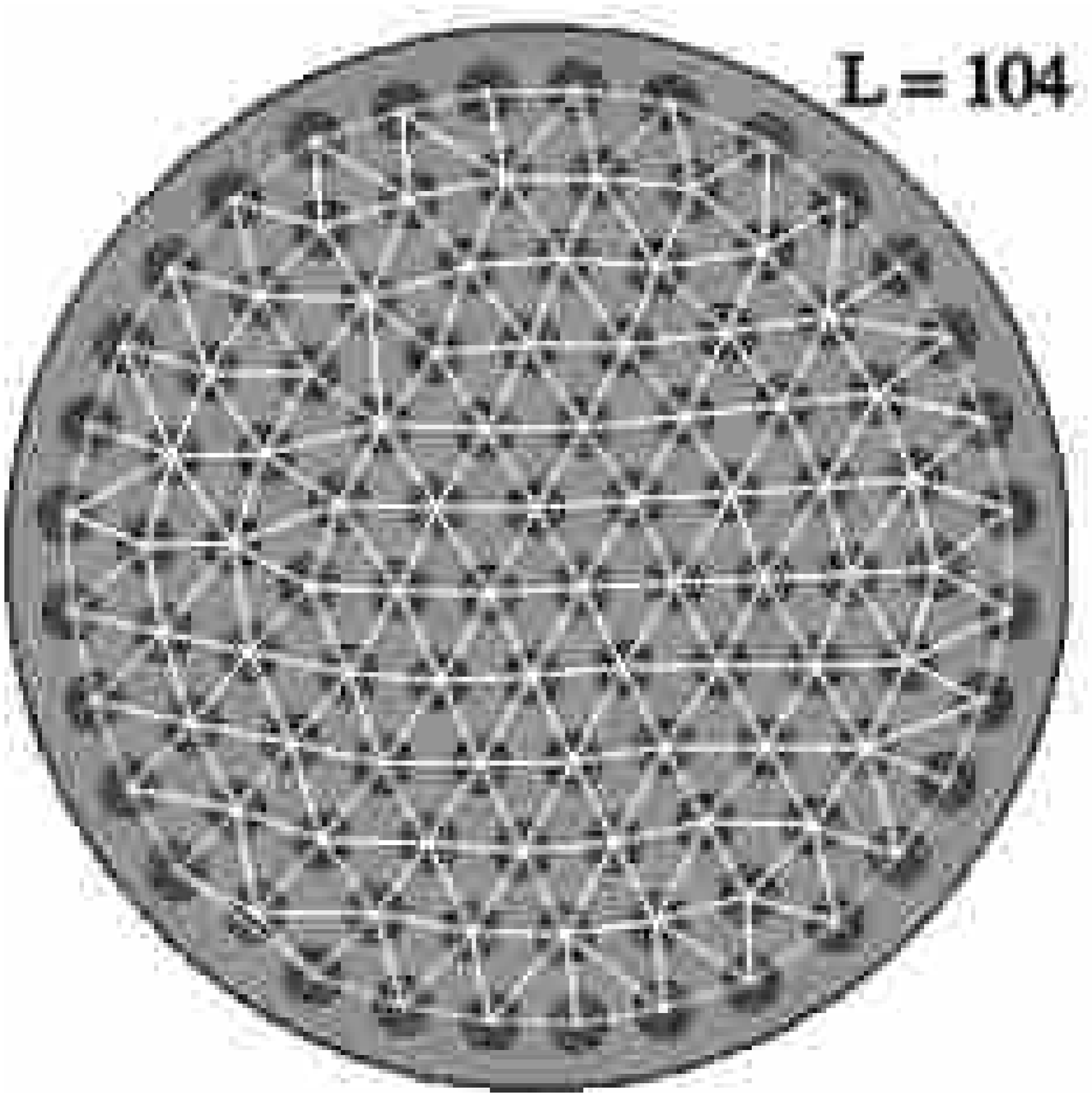}
\end{minipage}
\begin{minipage}[b]{0.47\linewidth}
\includegraphics[width=1\textwidth]{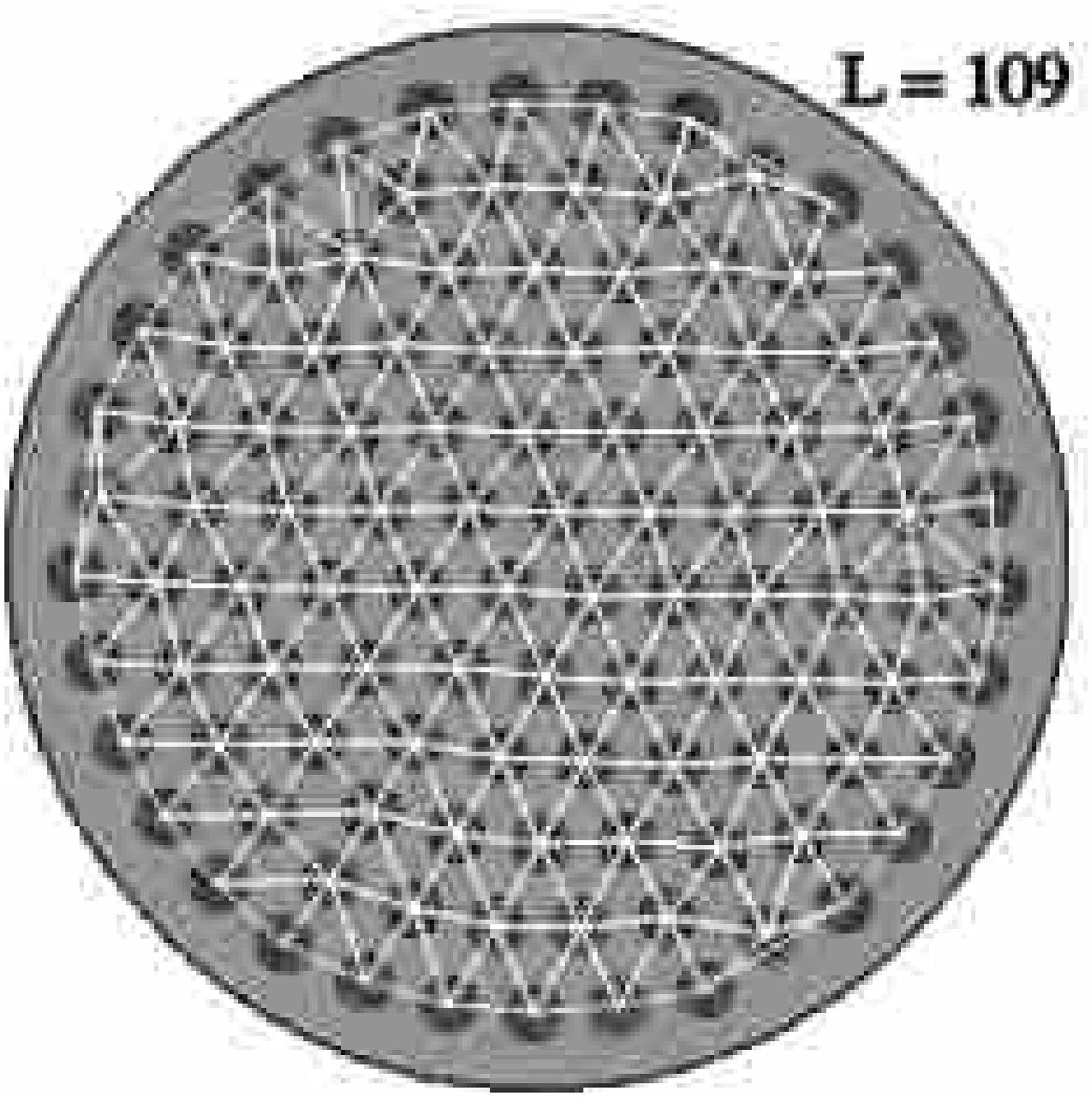}
\end{minipage}
\begin{minipage}[b]{0.47\linewidth}
\includegraphics[width=1\textwidth]{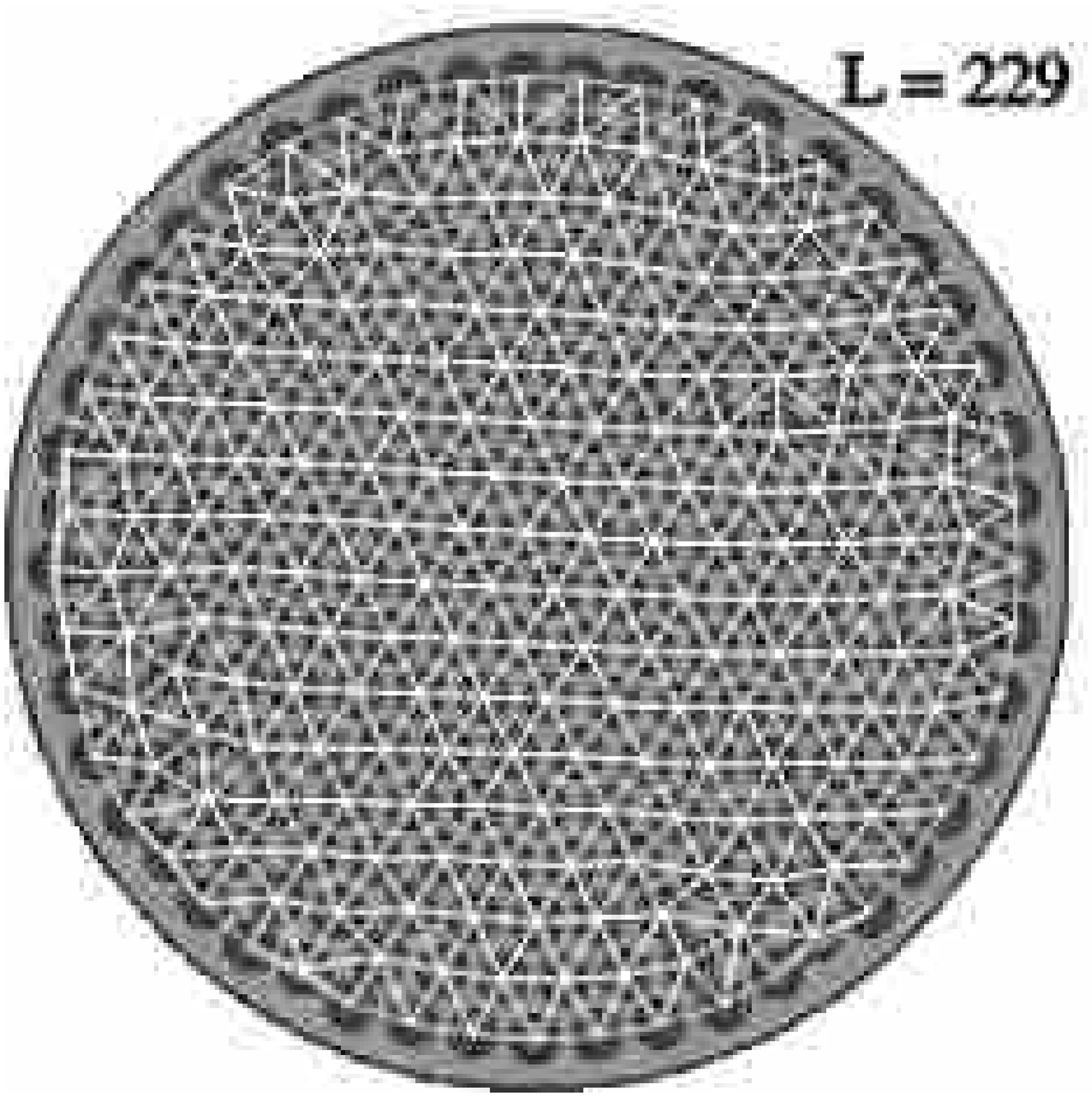}
\end{minipage}
 \vspace{0.25cm}
\caption{Superconducting electron density for $L = 44$, 56, 64,
79, 88, 104,  109, and 229 obtained at $H_0 = \,0.04$, 0.05, 0.06,
0.07, 0.08, 0.09, 0.10 and $0.20$, respectively. The white lines
depict the Delaunay triangulation for the vortex core
positions.\label{Fig_3} }
\end{figure}

In order to circumvent the limitations of the London limit in the
calculation of the energy, meta-state states are also investigated
within the GL theory. In this case, the correct contribution to
the energy from the spatial dependence of $|\Psi(\bm{r})|^2$ is
taken into account. Again, the question concerning whether the
calculated configurations are the true ground states can be
addressed, since it is possible that the numerical solution of
Eqs.~(\ref{lijn1},\ref{lijn2}) becomes trapped in some local
minimum. Nonetheless, thermal fluctuations are always present in
experiments, making some excited states close to the ground state
available for the system. In addition, there is the already
mentioned fact that the difference between energies in these high
$L-$states is very small. Therefore, the achievement of the ground
state is not crucial for the present study.

Although the London limit fails to give the precise value of the
vortex system energy at high $L$, we expect that the vortex
positions obtained within such approach are in good accordance
with the GL results (cf. Section~\ref{secii} and
Ref.~\onlinecite{Baelus_PRB03}), at least at fields up to $H_0
\sim 0.2$.~\cite{jltp26_735,Brandtreview} Therefore the stability
of the `London' configurations within the framework of the GL
theory was investigated by solving Eqs.~(\ref{lijn1})
and~(\ref{lijn2}) starting from the given London configuration
(usually the ones with lowest energy). By using this procedure, we
found that the  $L \sim 110$ and $L \sim 230$ configurations, as
obtained within the London theory, are also stable within the GL
formalism. The calculated GL energies of such configurations are
very close to other GL configurations with the same vorticity, the
relative difference lying typically between $10^{-4} - 10^{-5}$.
Such values are usually 5 to 10 times smaller than the relative
energy difference between the $L$ and $L+1$ lowest energy states.

Some of the stable configurations at $H_0 = \,0.04$, 0.05, 0.06,
0.07, 0.08, 0.09, 0.10 and $0.20$ are depicted in
Fig.~\ref{Fig_3}, for $L = 44$, 56, 64, 79, 88, 104,  109, and
229, respectively. From the Delaunay triangulation performed for
the core positions, it can be seen that a triangular vortex
configuration in the center of the disk starts to appear as $L$
increases. First, for $L = 64$ and $L = 79$, an hexagonal vortex
arrangement starts appearing in the center of the disk. Such
arrangement begins occupying a larger area with increasing
vorticity. For $L \gtrsim 100$ the Abrikosov lattice is already
present in a considerable region inside the disk.

\begin{figure*}[!tbp]
\centering
\begin{minipage}[t]{0.47\linewidth}
\includegraphics[width=1\textwidth]{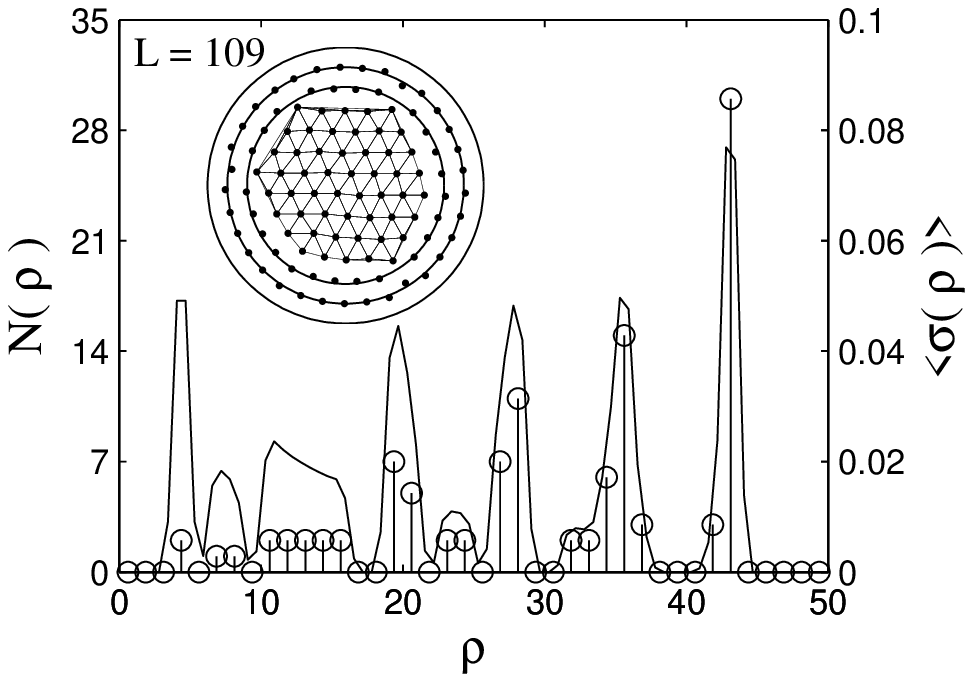}
\end{minipage}
\begin{minipage}[t]{0.47\linewidth}
\includegraphics[width=1\textwidth]{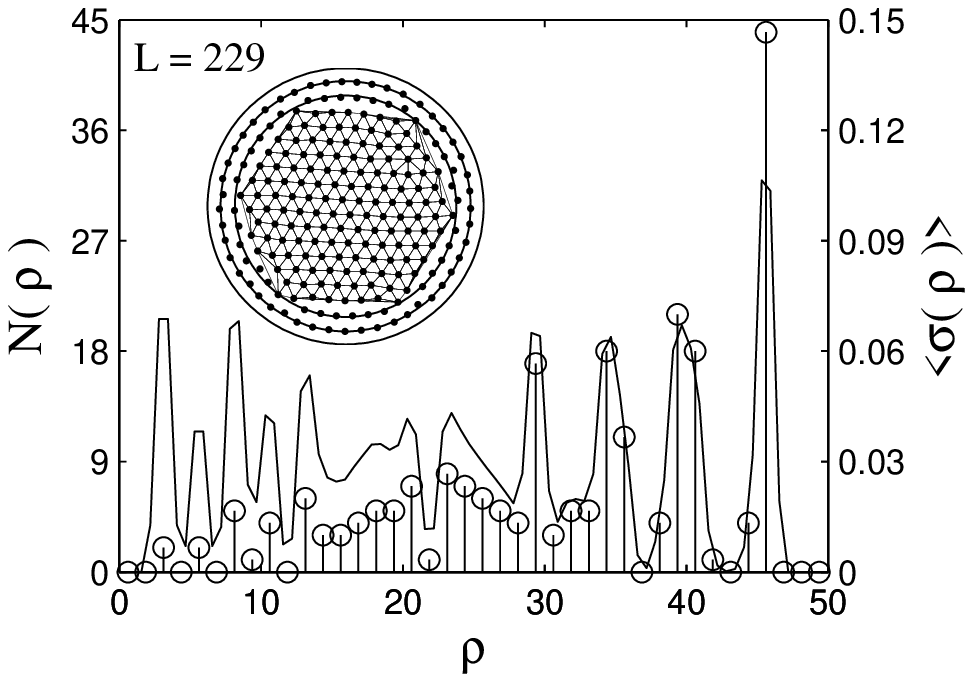}
\end{minipage}
\begin{minipage}[b]{0.47\linewidth}
\includegraphics[width=1\textwidth]{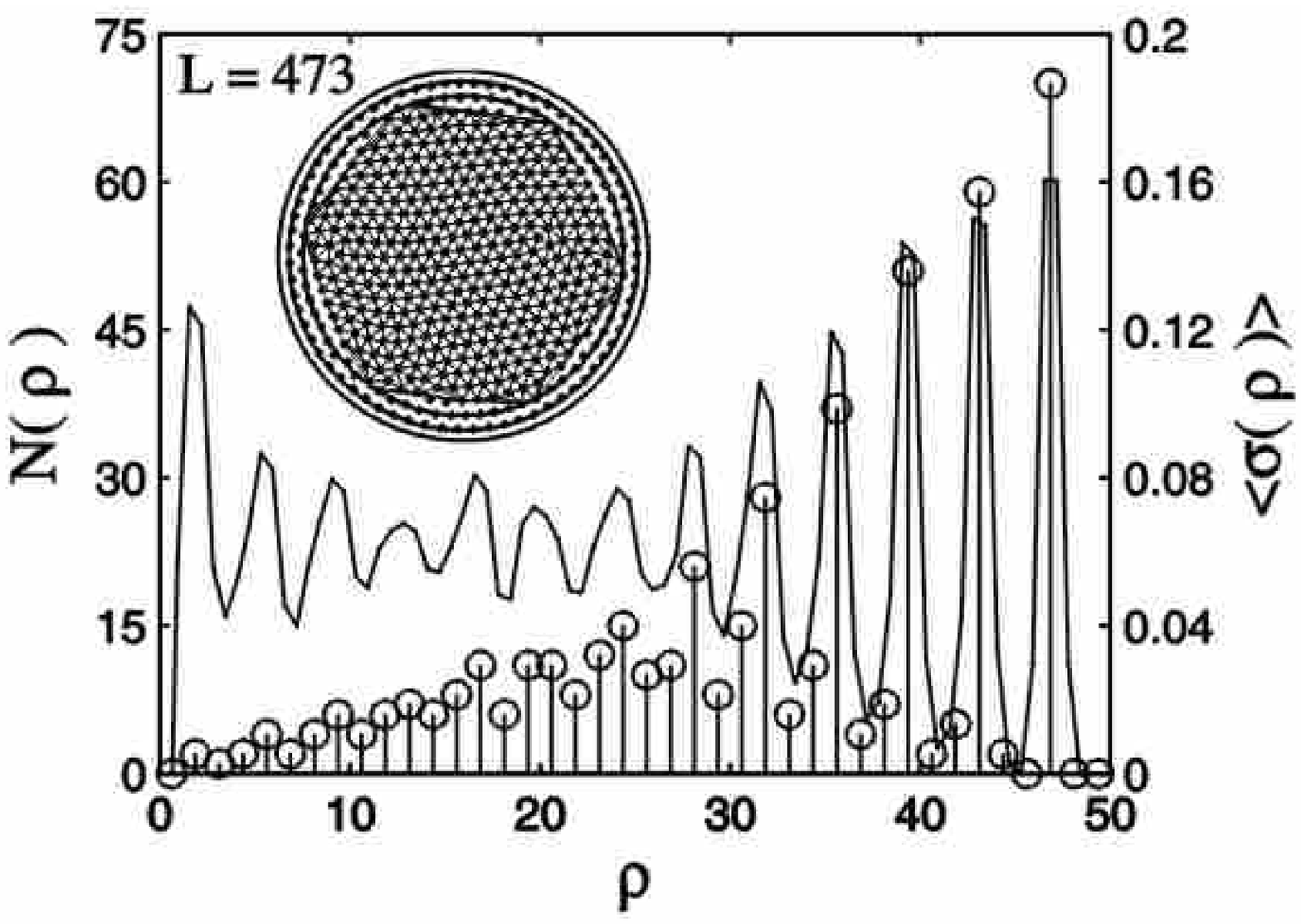}
\end{minipage}
\begin{minipage}[b]{0.47\linewidth}
\includegraphics[width=1\textwidth]{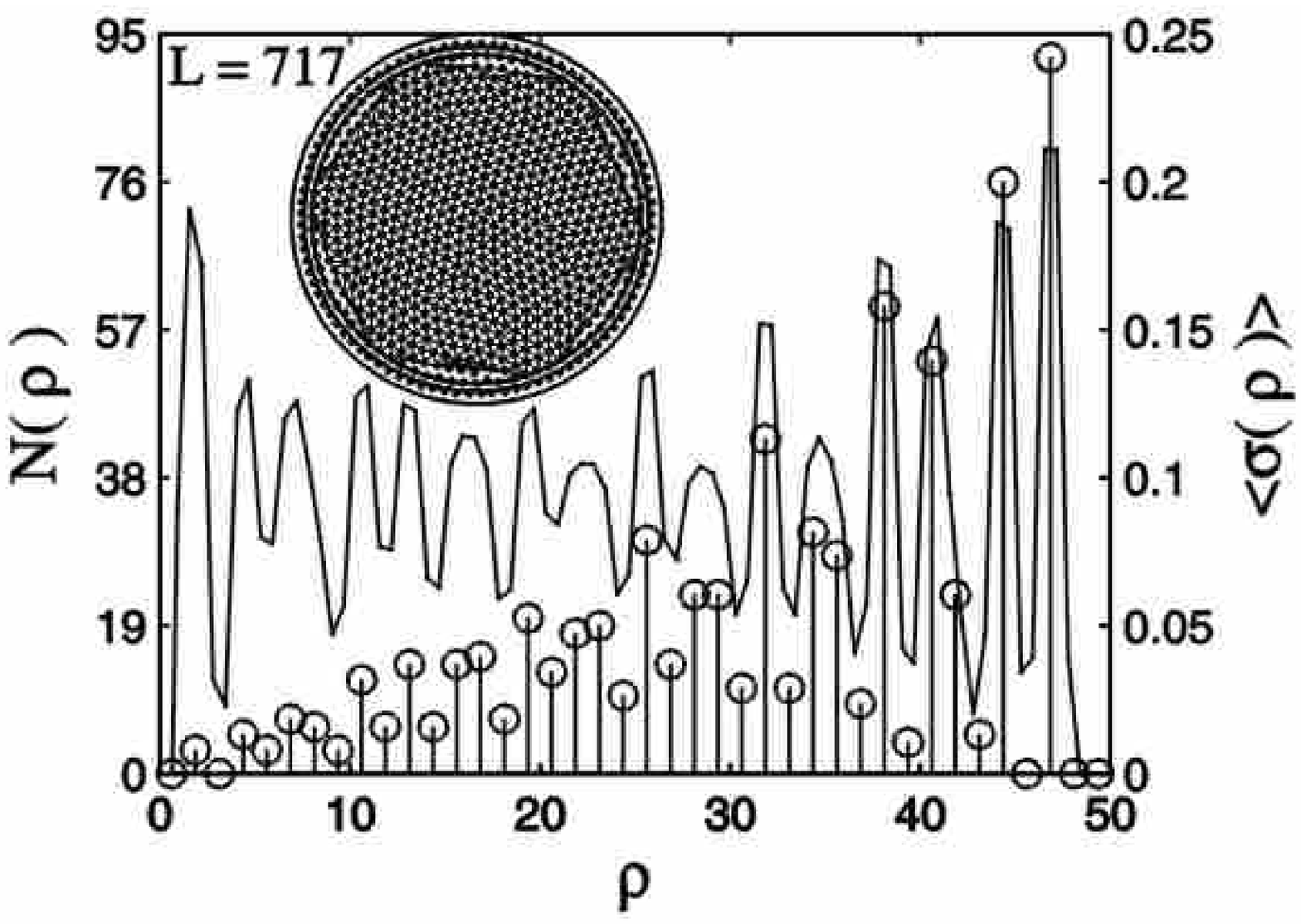}
\end{minipage}
 \vspace{0.0cm}
\caption{Number of vortices $N(\rho)$ (circles) and the average
vortex density $<\sigma(\rho)>$ (solid line) for $L = 109$, $L =
229$, $L = 473$ and $L = 717$ at, respectively, $H_0 = 0.1$, $H_0
= 0.2$, $H_0 = 0.4$ and $H_0 = 0.6$. The respective configurations
are depicted in the insets. The well-defined peaks close to $R =
50 \xi$ is indicative of a ringlike structure close to the edge.
This is also indicated by the configurations in the insets, where
we plotted rings for the two outermost shells and the Delaunay
triangulation for the inner vortices. \label{FigdensGL} }
\end{figure*}

For the high $L-$states there is a competition between the
ring-like structure imposed by the disk geometry, and the
hexagonal lattice favored by the vortex-vortex interaction. As a
result, rings are generally formed close to the disk edge while an
Abrikosov lattice is present in the center of the disk. In order
to study the configurations obtained within the GL theory, we
computed the positions of the vortex cores from the calculated
$\left|\Psi(\bm{r})\right|^2$.

\begin{figure*}[!tbp]
\centering
\begin{minipage}[t]{0.47\linewidth}
\includegraphics[width=1\textwidth]{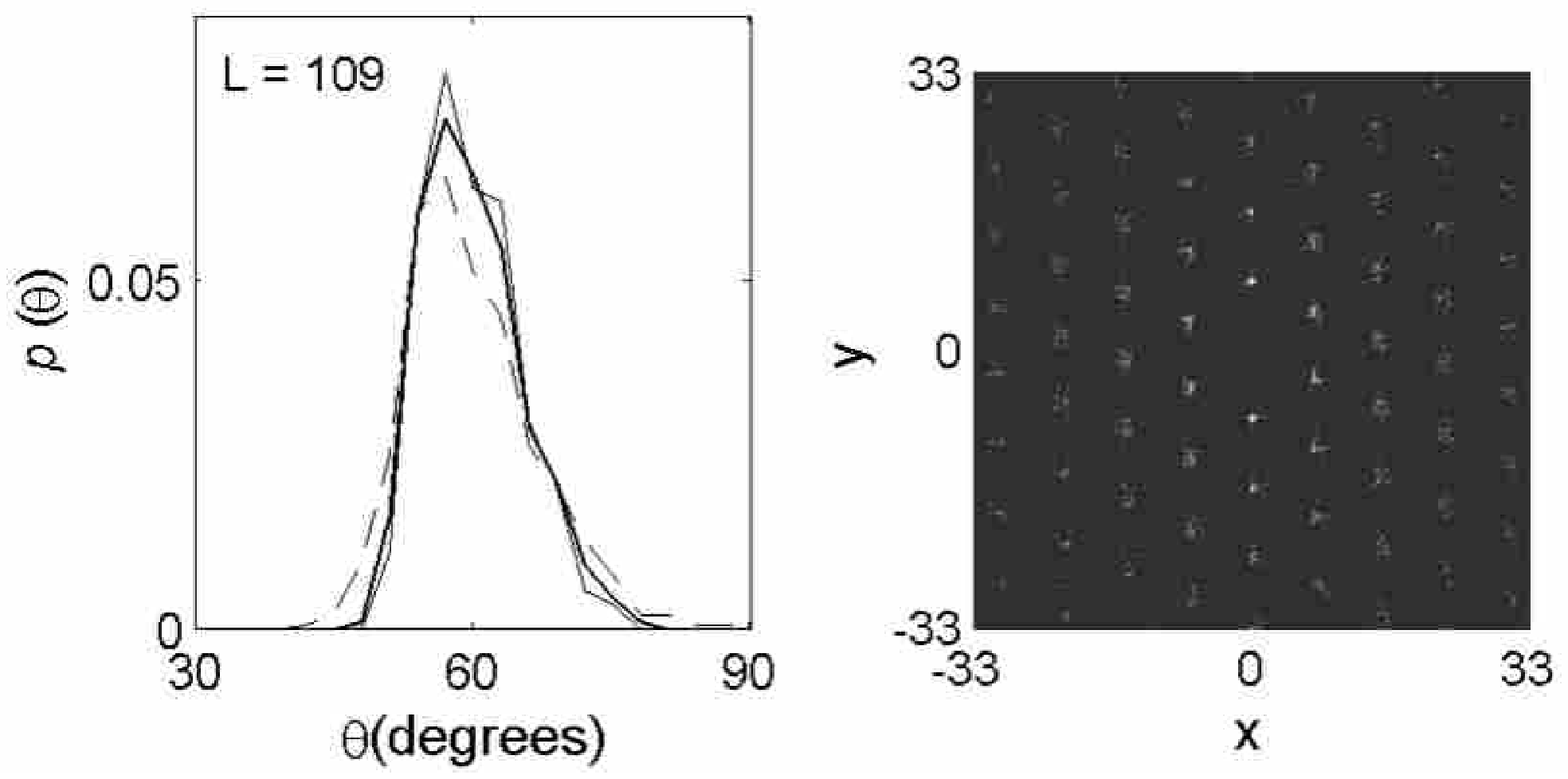}
\end{minipage}
\begin{minipage}[t]{0.47\linewidth}
\includegraphics[width=1\textwidth]{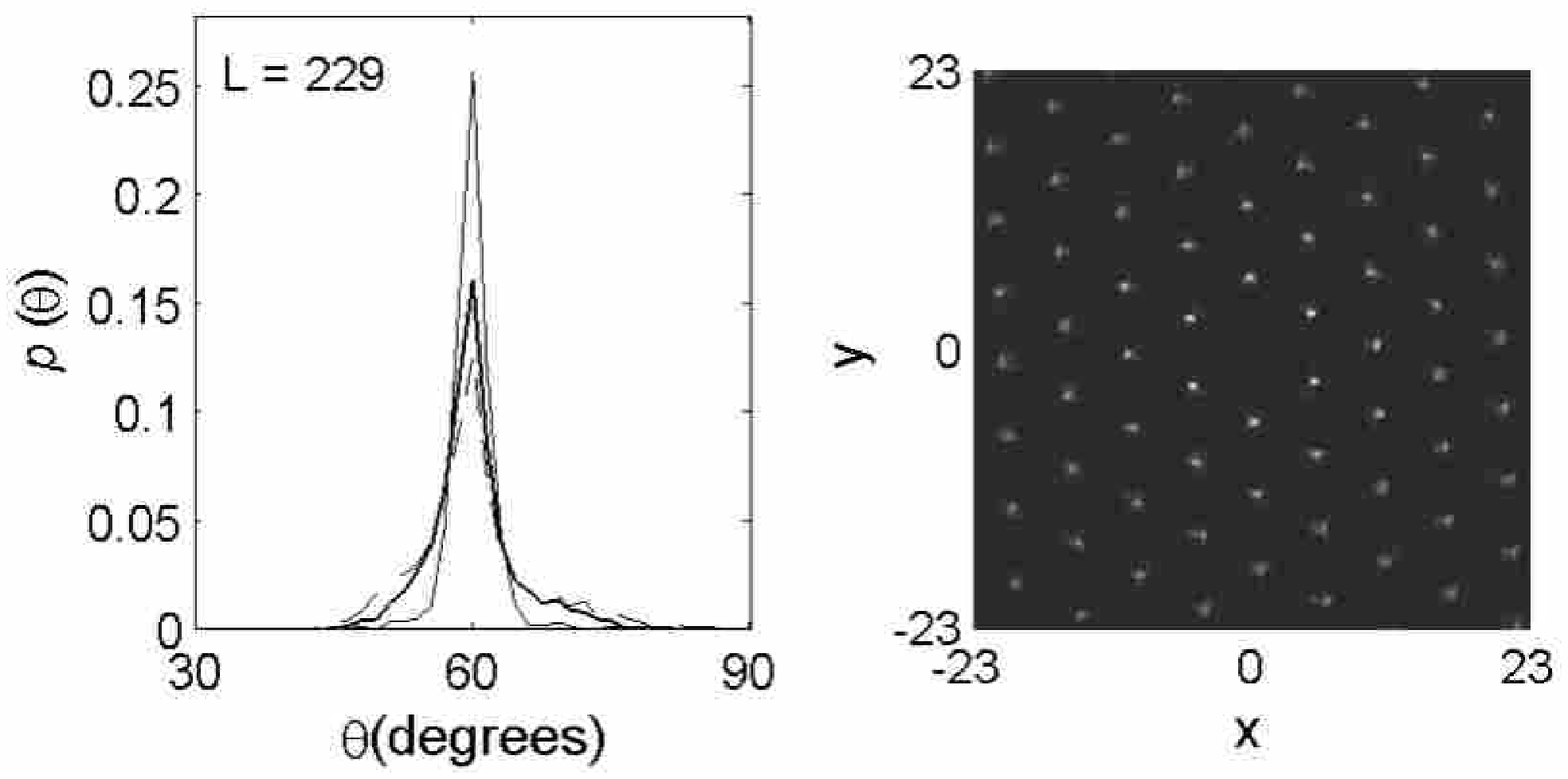}
\end{minipage}
\begin{minipage}[b]{0.47\linewidth}
\includegraphics[width=1\textwidth]{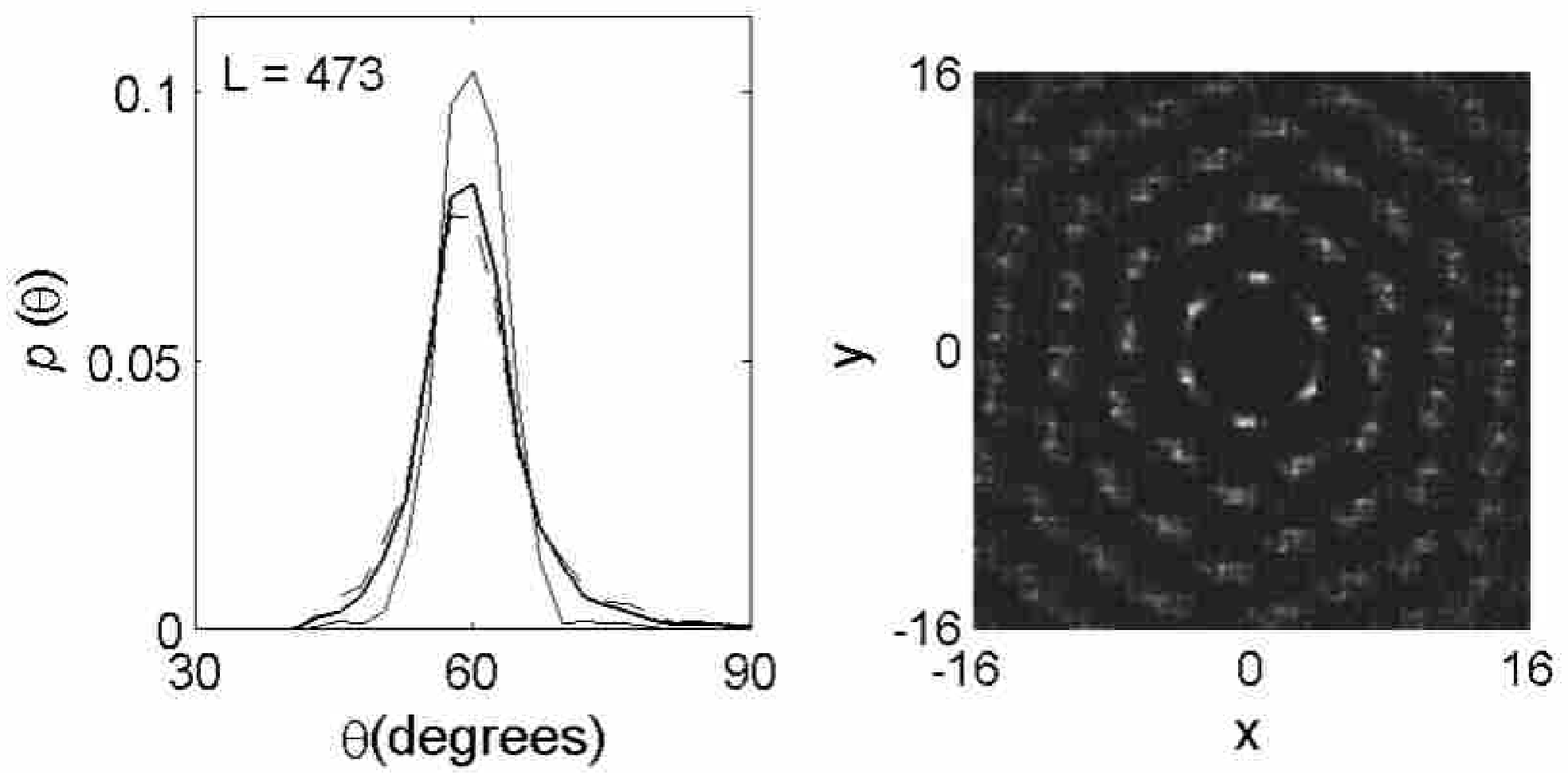}
\end{minipage}
\begin{minipage}[b]{0.47\linewidth}
\includegraphics[width=1\textwidth]{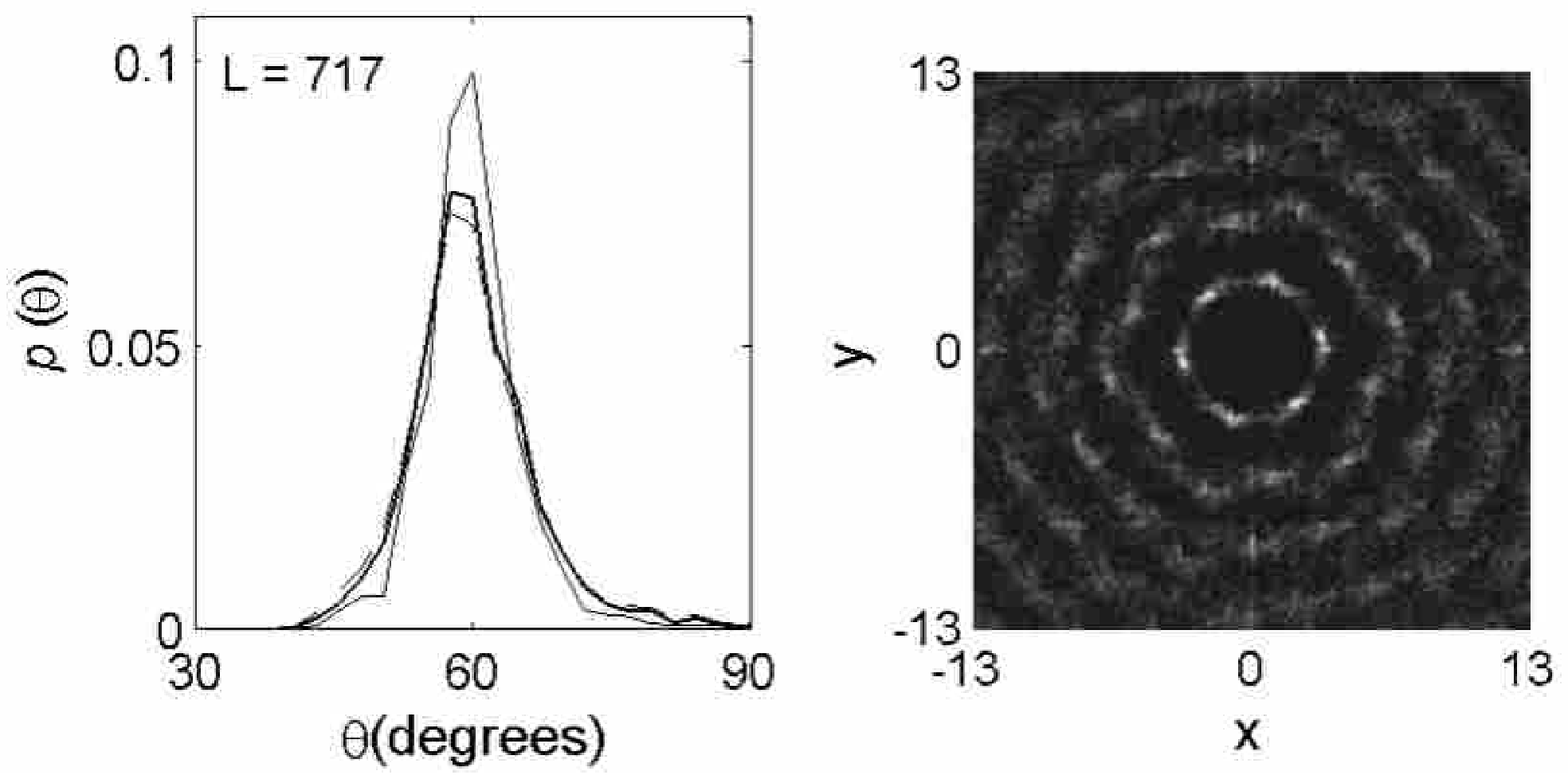}
\end{minipage}
 \vspace{0.0cm}
\caption{ The density-density correlation function (right) and the
probability $p(\theta)$ to find two adjacent nearest neighbors of
a given vortex within an angle $\theta$ (left) for $L = 109$ at
$H_0 = 0.1$, $L = 229$ at $H_0 = 0.2$, $L = 473$ at $H_0 = 0.4$,
and $L = 717$ at $H_0 = 0.6$.  The dashed, solid, and thin solid
lines represent $p(\theta)$ calculated for vortices $(i)$ not in
the outermost ring, $(ii)$ not in the two outer rings, and $(iii)$
at $\rho \le 25$, respectively. \label{FGLangdist} }
\end{figure*}
First we investigate the ringlike structure near the disk edge by
computing the number of vortices, $N$, and the average density of
vortices, $<\sigma(\rho)> = N(\rho)/2\pi\rho\Delta \rho$, as a
function of $\rho$. These quantities can suggest where ringlike
structures are formed, since $N(\rho)$ (as well as
$<\sigma(\rho)>$) should present sharp peaks where ringlike
patterns exist. For this purpose we divided the disk radius into
radial strips of length $\Delta \rho =1.25\xi$ and counted the
number of vortices in each of these pieces. $N(\rho)$ and
$<\sigma(\rho)>$ are shown in Fig.~\ref{FigdensGL} for $L = 109$,
$L = 229$, $L = 473$ and $L = 717$ at $H_0 = 0.1$, $H_0 = 0.2$,
$H_0 = 0.4$ and $H_0 = 0.6$, respectively. The $L = 109$ and $L =
229$ were obtained by solving the GL equations starting with the
$L = 110$ and $L = 230$ less energetic configurations calculated
within the London limit. We also plotted the respective
configurations inside each figure. To help the visualization,
rings were drawn for the two outermost shells and a Delaunay
triangulation was made for the vortices in the interior of these
rings. Clearly, both $N(\rho)$ and $<\sigma(\rho)>$ have one sharp
peak near the disk edge, an indication of a ring-like structure.
This can be observed in the vortex configurations since the
outermost vortices are almost perfectly aligned in a ring. For the
$L = 109$ state, both $N(\rho)$ and $<\sigma(\rho)>$ have
additional peaks in the interior of the disk. As the vortex
configuration also indicates, this could be interpreted as a
second (deformed) outer ring with a somewhat deformed hexagonal
lattice in the center. For $L = 229$, it is clear that vortices
are distributed in ring-like structures for the two outermost
rings with an inner Abrikosov lattice. Similar features are
present in the other $L \sim 110$ and $L \sim 230$ vortex states,
i.e., sharp peaks near the disk edge are also present in $N(\rho)$
and $<\sigma(\rho)>$, indicating two outermost ringlike vortex
distribution with an Abrikosov lattice in the center (again this
Abrikosov lattice is much better defined for $L \sim 230$). It is
also worth to mention that the two outer peaks present at $L \sim
110$ and $L \sim 230$ are situated around the same values of
$\rho$ for configurations calculated within both the GL and the
London theories. For example, for $L = 109$ the peaks are at $\rho
\approx 35$ and $\rho \approx 43$, with an empty region around
$\rho \approx 39$ and another for $\rho > 45$. Moreover the
regions comprised by the peaks in $<\sigma(\rho)>$ at $\rho
\approx 35$ and $\rho \approx 43$ contain $28$ and $33$ vortices,
respectively. In the case $L = 229$ one sharp peak occurs around
$\rho \approx 46$. The radial region close to this peak contains
$48$ vortices, with no vortices for $\rho > 47$. The radial region
around the peak at $\rho \approx 40$ has $44$ vortices, with the
region between these two maxima, around $\rho \approx 43$, also
vortex free. A more complete description of the number of vortices
in the two outer rings is presented in Table~\ref{table_1}. Taking
the number of vortices in the first and second outermost rings for
the configurations given in this Table, as well as other
configurations not shown here with the same vorticity, we find
that the number of vortices in these shells are around,
respectively, $33$ -- $34$ and $28\pm 1$ for $L \sim 110$ ($50\pm
2$ and $45 \pm 1$ for $L \sim 230$).

\begin{table}[tbp]
\caption{Number of vortices ($N$) and approximate radial position
of the two outer shells ($<\rho>$), and the bond-angular order
factor $G_6$ for configurations with lower energy. Here $(i)$
means all vortices, except the ones belonging to the outermost
shell; $(ii)$, vortices not at the two outer rings and $(iii)$
vortices at $\rho \le 25$.}
\begin{tabular}{lc|cc|cc|ccc}
\hline \hline \multicolumn{2}{c}{} & \multicolumn{2}{c}{1st.
shell} & \multicolumn{2}{c}{2nd. shell} &
\multicolumn{3}{c}{$G_6$}
\\
 $L$ & $H_0$ & $N$ & $<\rho>$ & $N$ & $<\rho>$ & $(i)$ &
$(ii)$ & $(iii)$
\\ \hline
$109$ & $0.1$ & 33 & $43$ & 28 & $35$ & $0.76$ & $0.85$ & $0.87$ \\
$110$ & $0.1$ & 33 & $43$ & 28 & $35$ & $0.64$ & $0.71$ & $0.75$ \\
$111$ & $0.1$ & 33 & $43$ & 29 & $36$ & $0.69$ & $0.79$ & $0.84$ \\
$112$ & $0.1$ & 33 & $43$ & 29 & $36$ & $0.68$ & $0.80$ & $0.88$ \\
$113$ & $0.1$ & 34 & $43$ & 28 & $36$ & $0.70$ & $0.80$ & $0.84$ \\
$229$ & $0.2$ & 48 & $46$ & 44 & $40$ & $0.80$ & $0.89$ & $0.97$ \\
$230$ & $0.2$ & 48 & $46$ & 44 & $40$ & $0.78$ & $0.84$ & $0.94$ \\
$231$ & $0.2$ & 50 & $46$ & 44 & $40$ & $0.83$ & $0.92$ & $0.99$ \\
$232$ & $0.2$ & 49 & $46$ & 44 & $40$ & $0.82$ & $0.91$ & $0.97$ \\
$233$ & $0.2$ & 49 & $46$ & 45 & $41$ & $0.80$ & $0.87$ & $0.96$ \\
$234$ & $0.2$ & 50 & $46$ & 45 & $41$ & $0.81$ & $0.87$ & $0.95$ \\
$235$ & $0.2$ & 49 & $46$ & 44 & $41$ & $0.82$ & $0.90$ & $0.97$ \\
$473$ & $0.4$ & 70 & $47$ & 66 & $43$ & $0.79$ & $0.83$ & $0.92$ \\
$717$ & $0.6$ & 92 & $47$ & 80 & $44$ & $0.77$ & $0.79$ & $0.86$ \\
\end{tabular}
\label{table_1}
\end{table}
In Fig.~\ref{FigdensGL} the states $L=473$, at $H_0 = 0.4$ and
$L=717$, at $H_0 = 0.6$, are also depicted. As expected, the peaks
become broader deep inside the disk, suggesting that the ring-like
structure smears out as one approaches the center of the disk. In
addition, as the value of $L$ increases the average density
becomes more uniform, but preserving at least two sharp peaks near
the edge. For $L = 473$ and $L = 717$ the most external ring is
situated at $\rho \approx 47$ and contains 70 and 92 vortices,
respectively. Notice that the two outer rings have a very
different number of vortices which is quite distinct from the
situation of classical charges confined by a parabolic
potential~\cite{PRB49b} where for large number of charges the
outer rings contain the same number of particles, The present
situation is between a hard wall~\cite{KongHW} and a parabolic
confinement case.

We calculated the density-density correlation function for the
vortices sitting inside the two outermost rings in order to help
characterize whether a Abrikosov lattice is formed away from the
disk edge. This quantity is depicted at the right side of
Fig.~\ref{FGLangdist}. The density-density correlation function
indicates an hexagonal pattern for all these high $L-$states. Such
pattern is well defined for $L = 109$ at $H_0 = 0.1$ and becomes
very well defined for $L = 229$ at $H_0 = 0.2$. Other
configurations with $L \sim 110$ have also an hexagonal pattern as
the one for $L = 109$ (but not as sharp). The density-density
correlation function computed for various configurations with $L
\sim 230$ also resembles the one depicted here for $L = 229$. For
$L = 473$ and $L = 717$ the hexagonal pattern is also observed,
but not as sharp as the one for $L = 229$. Particularly for the $L
= 717$ configuration, the density-density correlation function
suggests that each vortex (inside the two outermost rings) still
have coordination number equals to six, although the hexagonal
structure considering the farther neighbors is not well defined.
Therefore these two configurations may still have local, but not
orientational order beyond some few neighbors. We shall come back
to this point later in Section~\ref{secv}, when discussing the
defects in the vortex lattice.

From the density-density correlation function it is also worth to
compute the typical inter-vortex distance, $a_v$, for the vortices
forming the Abrikosov lattice. We thus obtained $a_v \approx 8$
for $L = 109$ at $H_0 = 0.1$, $a_v \approx 5.8$ for $L = 229$ at
$H_0 = 0.2$, $a_v \approx 4.1$ for $L = 473$ at $H_0 = 0.4$, and
$a_v \approx 3.4$ for $L = 717$ at $H_0 = 0.6$.

In order to better describe how close the system is to an
Abrikosov lattice we computed the probability distribution,
$p(\theta)$, to find two adjacent nearest neighbors of a given
vortex making an angle $\theta$. This probability was calculated
for three different cases: ($i$) for all vortices, except the ones
at the outermost ring; ($ii$) for the vortices not in the two
outer rings, and ($iii$) for those vortices at $\rho \le 25$.
These probabilities are shown on the left of
Figs.~\ref{FGLangdist}. We found that $p(\theta)$ (for all the
cases $(i) \to (iii)$) is maximum close to $60^{\rm o}$, which is
characteristic of an hexagonal lattice. The width of the
distribution rapidly decreases as $L $ increases from $\sim 110$
to $\sim 230$, but increases as $L$ is further incremented. To be
more precise, $p(\theta)$ for the $L \sim 110$ (not only the $L =
109$ state which is shown) state obtained within the London limit
has a maximum at $57^{\rm o}$, but with $<\theta> = 60^{\rm o}$
for the cases $(i) \to (iii)$ ($<>$ means average over the
vortices included in each case, $(i)$, $(ii)$ or $(iii)$). The
probability distributions for cases $(i) \to (iii)$ are not sharp,
presenting width of about $12^{\rm o}$ at half of the distribution
maximum. Other states with $L \sim 110$ and comparable energy also
show similar behavior. Such features can be understood as the
result of the contribution to the $p(\theta)$ distribution from
vortices in the border of the Abrikosov lattice region. Since not
so many vortices are present in this region for $L \sim 110$,
vortices in its border will contribute more strongly to the
$p(\theta)$ distribution than for higher $L$ states. Such vortices
have to adjust themselves to the ring-like structure more than the
inner vortices and, so, it is likely that a few of them may have
nearest neighbors within angles less than $60^{\rm o}$ or, even,
coordination number different to six.  For $L > 200$, $p(\theta)$
is sharply peaked at $\theta = 60^{\rm o}$, in conformity with the
density-density correlation function, signaling an Abrikosov
lattice in the interior of the disk.

For completeness we also calculate the bond-angular order
factor~\cite{HalperinNelson,PRL82p5293},
\begin{eqnarray}
G_6 =
\left<\frac{1}{N_{nb}}\sum_{n=1}^{N_{nb}}\exp(iN_{nb}\theta_n)\right>,
\label{eq_L12}
\end{eqnarray}
where $N_{nb} = 6$ is the number of nearest neighbors of a given
vortex, $\theta_n$ is the angle between two segments joining the
given vortex with two adjacent nearest neighbors and $<>$ is again
the average over the vortices in cases $(i)$, $(ii)$ or $(iii)$.
It is clear from Eq.~(\ref{eq_L12}) that $G_6 = 1$ for an ideal
Abrikosov lattice. In table~\ref{table_1} $G_6$ is depicted for
some of the configurations we obtained (typically the
configurations with lowest energy). The values found for $G_6$ are
larger than $0.9$ at the region $(iii)$ for $L \sim 230$, which
indicates a configuration very close to an hexagonal lattice. The
$L \sim 110$ states obtained at $H_0 = 0.1$ have lower $G_6$,
which corroborates our previous analysis suggesting that an
Abrikosov lattice is formed but not yet occupying a large area
inside the disk. Again, for $L = 473$ and $L = 717$ $G_6$ is not
as large as the one calculated at $L \sim 230$, but is still close
or larger than $0.9$ in region (iii), which indicates that a local
orientational hexagonal order is present. In fact for such large
$L-$values $G_6$ no longer increases and the peak in $p(\theta)$
is slightly broadened due to the appearance of grain boundaries in
the Abrikosov lattice as will be shown in the next section.

\section{High $L-$states: Defects in the Vortex Lattice}
\label{secv}

\begin{figure*}[!tbp]
\centering
\begin{minipage}[t]{0.3\linewidth}
\includegraphics[width=1\textwidth]{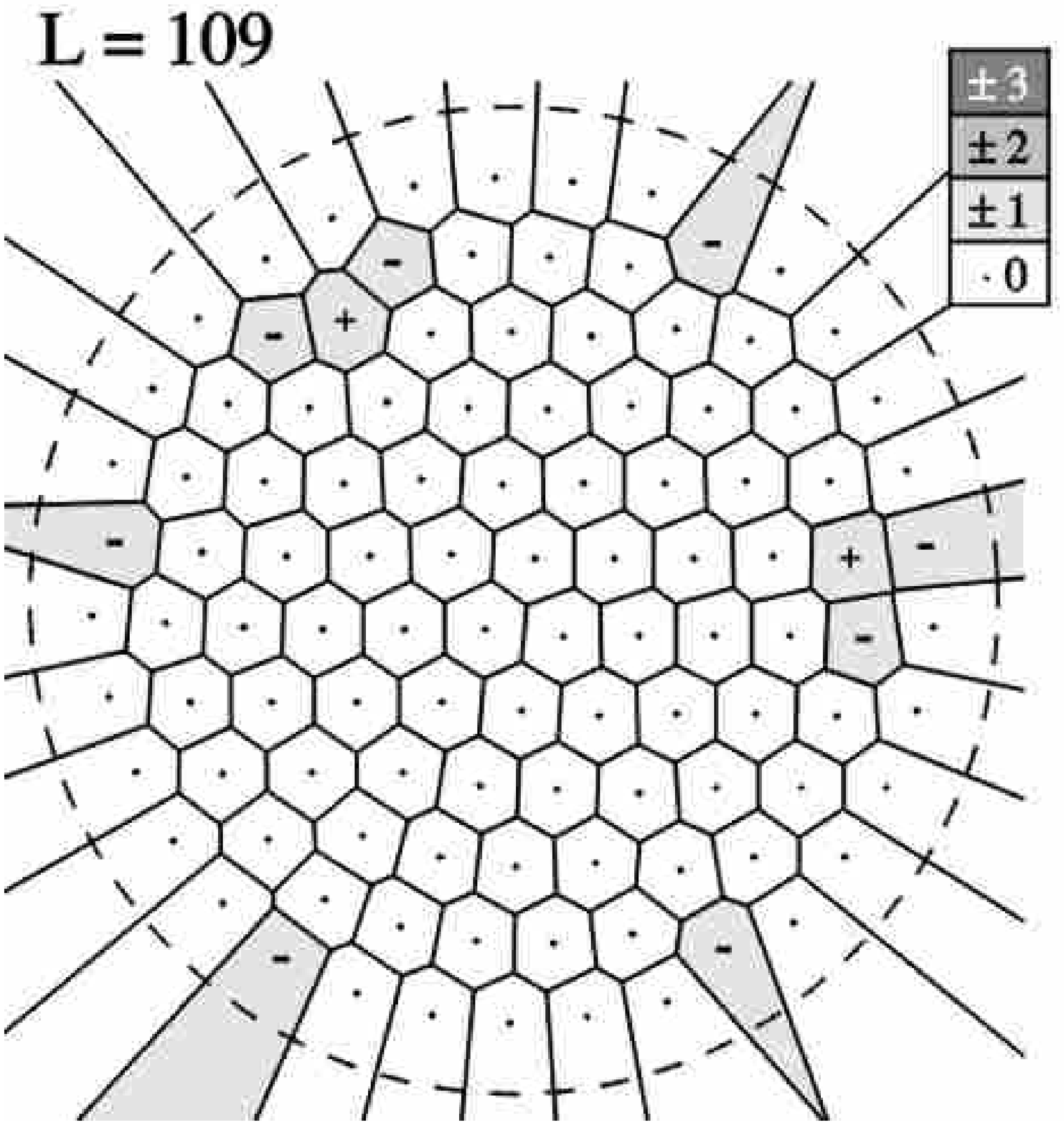}
\end{minipage}
\begin{minipage}[t]{0.3\linewidth}
\includegraphics[width=1\textwidth]{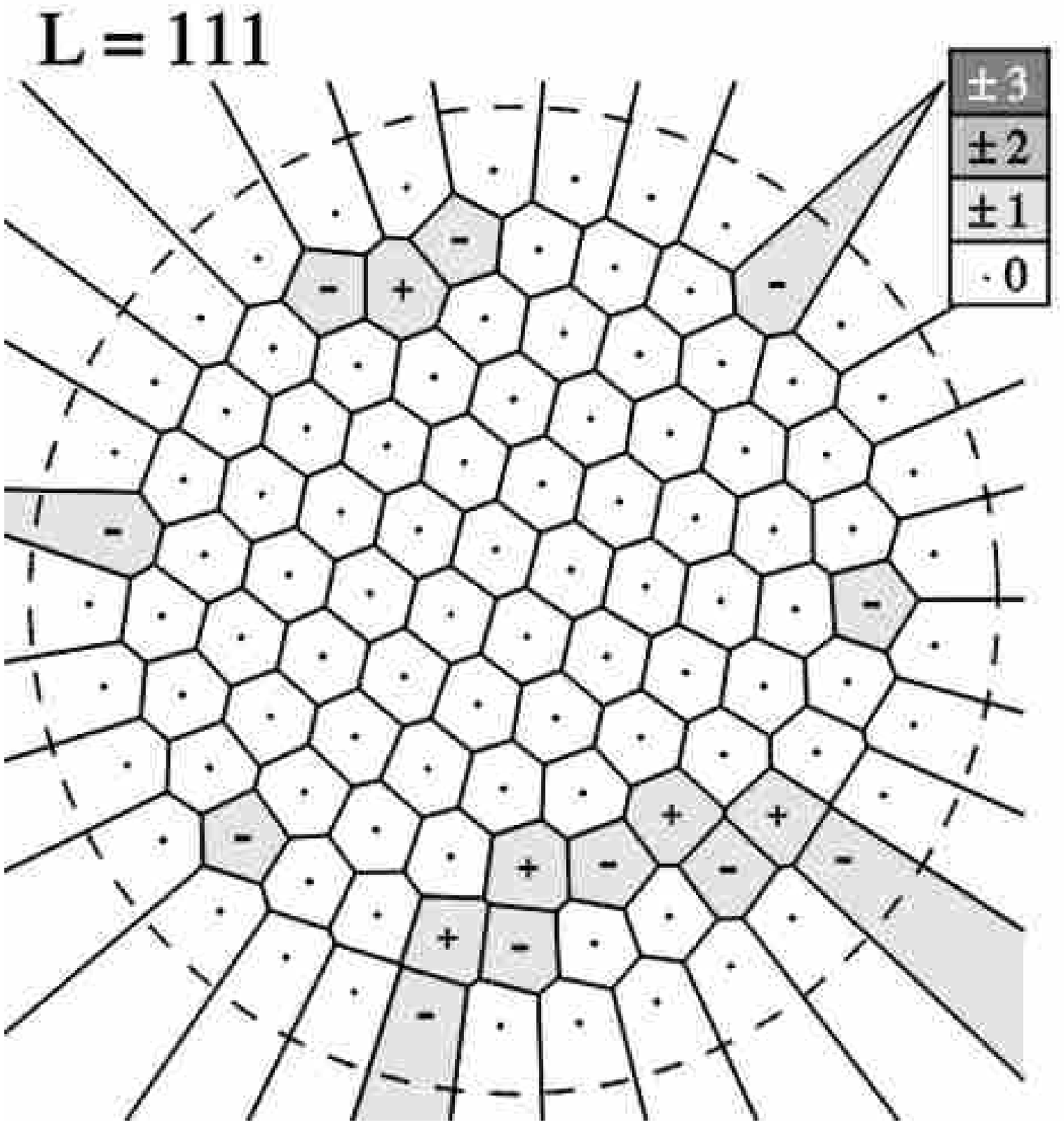}
\end{minipage}
\begin{minipage}[t]{0.3\linewidth}
\includegraphics[width=1\textwidth]{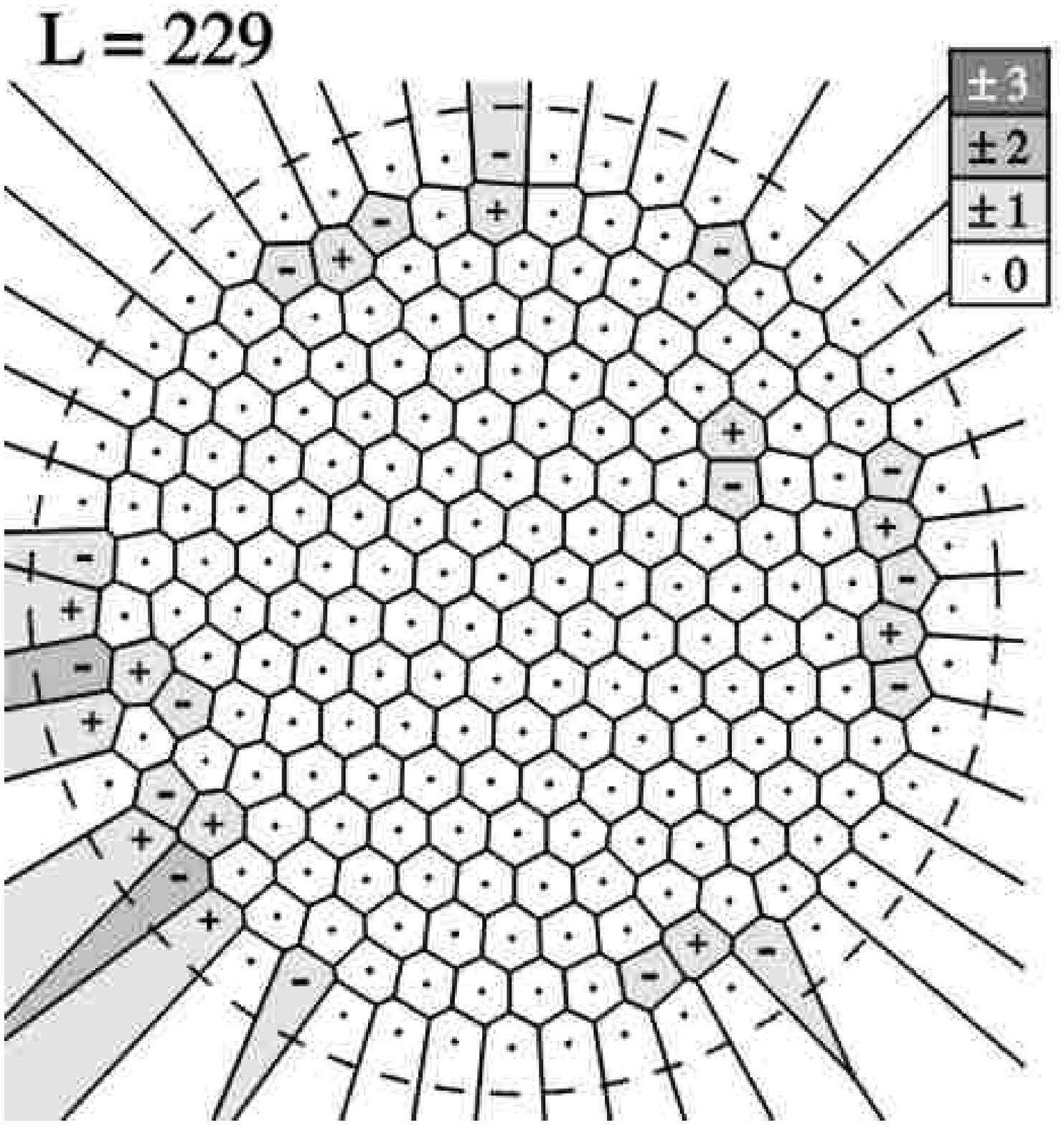}
\end{minipage}
\begin{minipage}[b]{0.3\linewidth}
\includegraphics[width=1\textwidth]{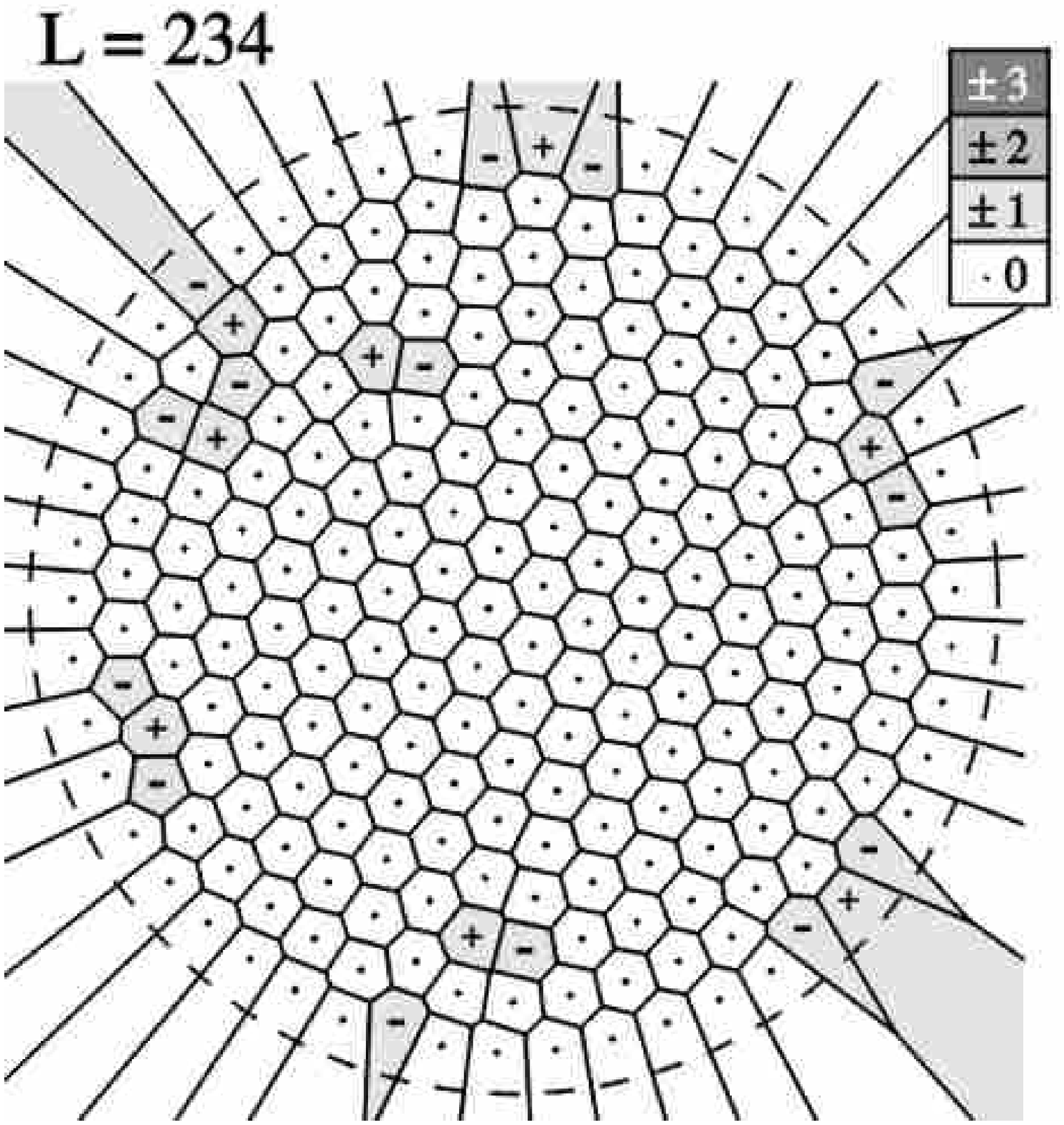}
\end{minipage}
\begin{minipage}[b]{0.3\linewidth}
\includegraphics[width=1\textwidth]{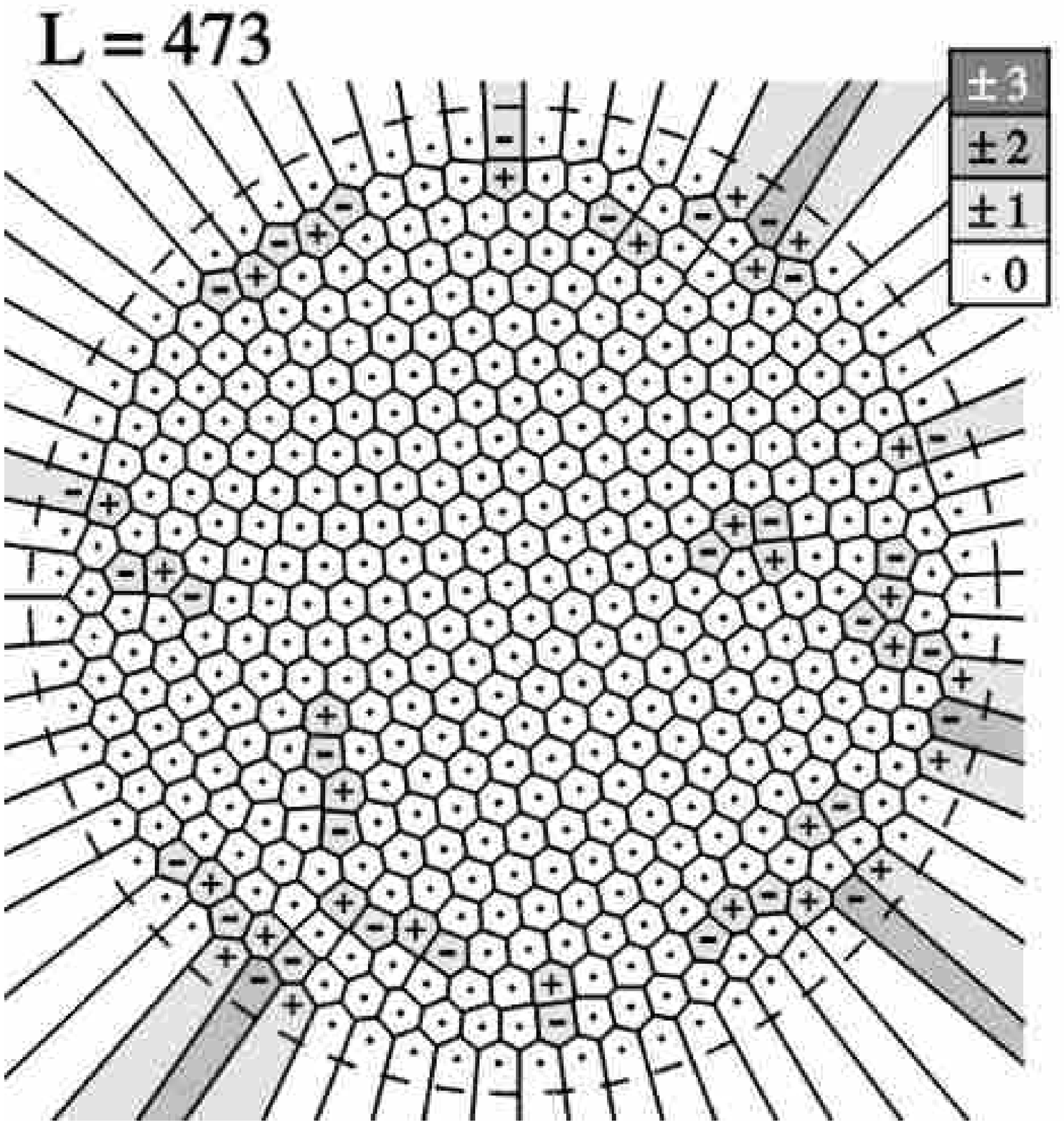}
\end{minipage}
\begin{minipage}[b]{0.3\linewidth}
\includegraphics[width=1\textwidth]{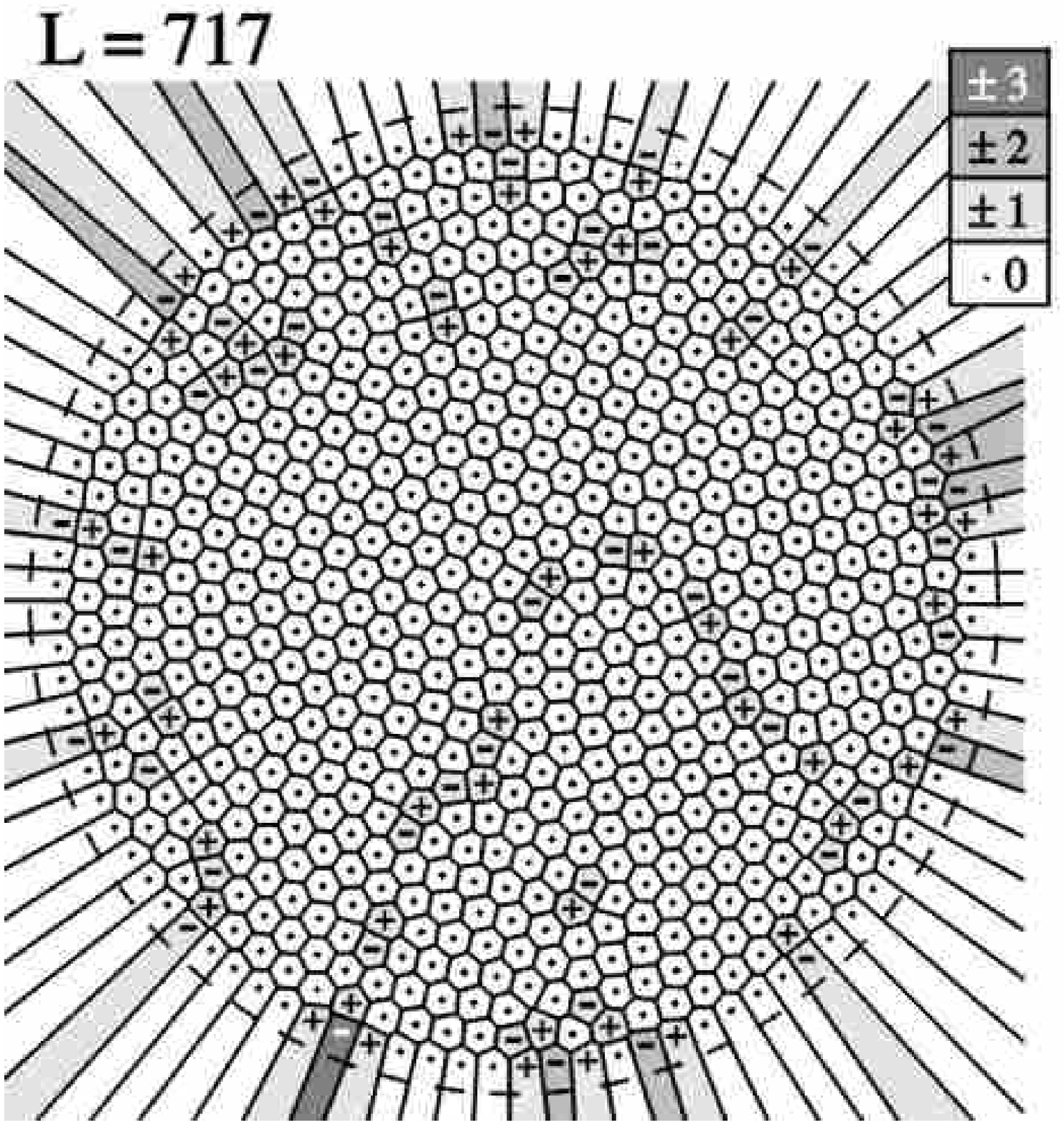}
\end{minipage}
 \vspace{0.0cm}
\caption{ Voronoi construction for the following configurations:
$L = 109$ and $L = 111$ at $H_0 = 0.1$, $L = 229$ and $L = 234$ at
$H_0 = 0.2$, $L = 473$ at $H_0 = 0.4$, and $L = 717$ at $H_0 =
0.6$. The dashed line represents the disk edge.\label{VorGL} }
\end{figure*}
As a result of the competition between the geometry induced
ring-like structure near the disk border and the hexagonal
structure in the center, topological defects in the lattice appear
in between these two regions (a feature also observed in confined
classical systems~\cite{PRB57_2352,pre67_021608}). In order to
study the distribution of these deffects in the disk, we applied
the Voronoi construction. In an infinite system both the GL theory
and the London approach predict a coordination number equal to six
and the Voronoi construction would yield hexagonal unit cells for
each vortex. In the disk the situation is different, vortices near
the edge have to adjust themselves to the boundary. Therefore,
topological defects in the vortex lattice will be present. We
shall use the term (wedge) disclination for vortices which have a
closed unit cell in the Voronoi construction with coordination
number different from six. This difference is called the
topological charge of the disclination. Notice that some vortices
at the outermost shell have open unit cells in the Voronoi
construction. For such vortices the expected number of nearest
neighbors should be four. So in order to define the topological
defects also for these vortices, the topological charge there is
defined as the number of first neighbors minus $4$. By such
convention it can be shown from Euler's theorem~\cite{PRB57_2352}
that the net topological charge in a disk equals $-6$. In
addition, dislocations (a bounded pair of one $+$ and one $-$
disclinations) may also appear, whose net topological charge is
null, in order to adjust the vortex system to a configuration with
lower energy.

Fig.~\ref{VorGL} shows the Voronoi construction for the $L = 109$
($H_0 = 0.1$), $L = 111$ ($H_0 = 0.1$), $L = 234$ ($H_0 = 0.2$),
$L = 229$ ($H_0 = 0.2$), $L = 473$ ($H_0 = 0.4$), and $L = 717$
($H_0 = 0.6$). In all of them it is quite clear that an Abrikosov
vortex lattice is formed inside the disk, as indicated in previous
section, but with the formation of topological defects in the
configurations. The net topological charge for all configurations
obtained (including the ones not shown here) is always $-6$, in
accordance with the Euler theorem.~\cite{PRB57_2352} However the
total absolute charge can be much larger than $6$. Negatively
charged disclinations (vortices with coordination number $< 6$)
are always present. Vortices with coordination number $> 6$
(positive topological charge) appear accompanied by negative
topological charges, leading to the formation of dislocations. The
defects in the vortex configurations are more suitable to sit in
the disk edge or in the region delimiting the Abrikosov lattice
and the ring-like structure. Nevertheless, as $L$ increases,
dislocations proliferate and form grain boundaries in the region
where the hexagonal lattice appears. This is also the reason why
the $L = 473$ and $L = 717$ states have smaller $G_6$ values and
less sharper peaks in the $p(\theta)$ distribution than the lower
$L$ states, for instance $L \sim 230$. Such feature is also
observed in simulations performed by Reefman and
Brom~\cite{PhysC183_212} considering $2000$ vortices (although
they considered vortices in the London limit without interaction
with the disk edge) and in classical systems of charged particles
interacting with each other via the Coulomb potential and confined
to a parabolic potential.~\cite{pre67_021608}

Koulakov and Shklovskii~\cite{PRB57_2352} described the presence
of dislocations in configurations of classical charged particles
confined by a parabolic potential as due to two main reasons: the
inhomogeneity in the density of particles and the presence of
disclinations. The latter (which is always present in an hexagonal
arrangement confined to a disk) causes a large deformation in the
particle configurations. Dislocations thus appear in order to
reduce such deformations, eventually decreasing the energy of the
system. Such effect, also called screening, was previously
described by Nelson and Halperin~\cite{HalperinNelson} when
studying the melting driven by dislocations in two dimensional
systems, and is linked to the lack of translational long-range
order in two-dimensional solid systems (although orientational
order is still present).~\cite{NDMermin} These dislocations are
arranged close to or at the disk edge. The former reason induces
dislocations in the interior of the disk. In
Ref.~\onlinecite{PRB57_2352}, it was found that there exists a
threshold number of particles (which in their case is
approximately $700$) below which dislocations are due mainly from
screening and, above which, such defects appear due to the
inhomogeneity of the particle density. At least qualitative
similarities exists between such system of charged particles and
our vortex configurations. Therefore, it is reasonable to
speculate that the same mechanisms which drive the appearance of
dislocations is also present here. Just like in the  system of
charged particles, dislocations are mostly distributed close to
and at the disk edge for $L\lesssim 230$ and start proliferating
in the Abrikosov lattice for larger $L$.
\begin{figure}[!tbp]
\centering
\begin{minipage}[t]{0.49\linewidth}
\includegraphics[width=1\textwidth,,height=1.43\textwidth]{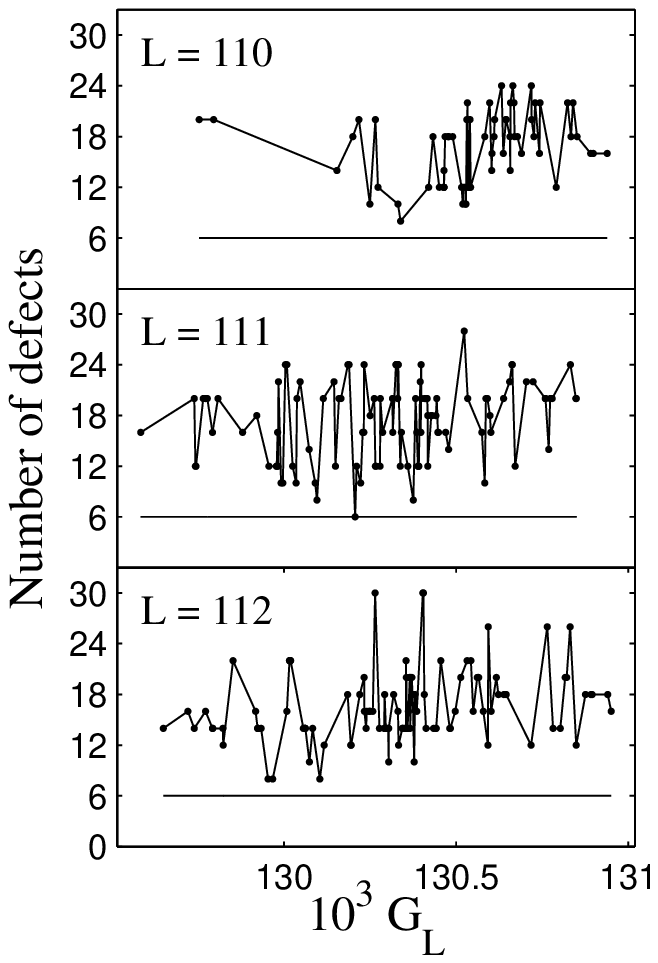}
\end{minipage}
\begin{minipage}[b]{0.49\linewidth}
\includegraphics[width=1\textwidth,height=1.43\textwidth]{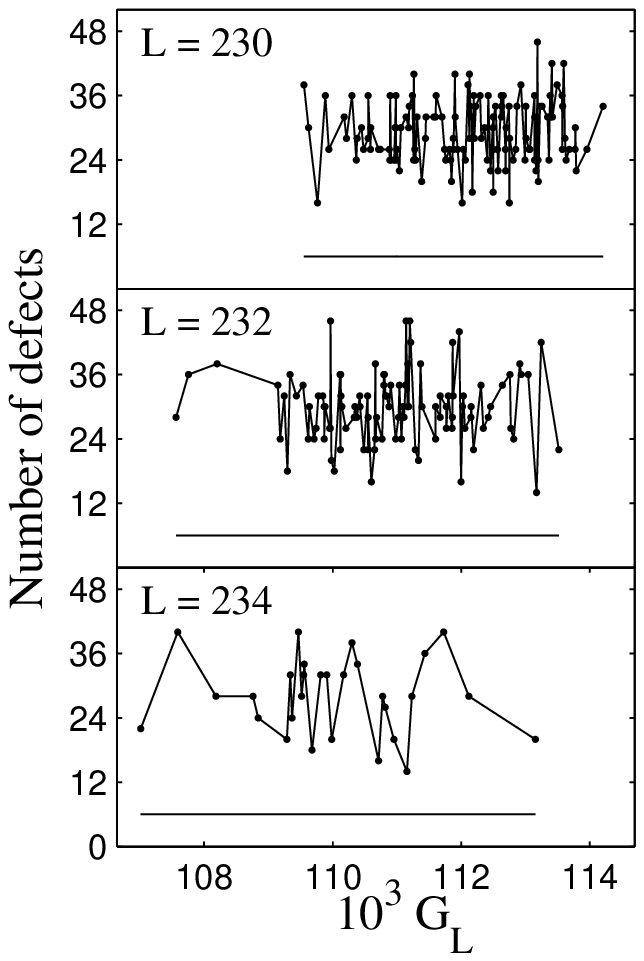}
\end{minipage}
 \vspace{0.0cm}
\caption{ Number of defects (solid points) versus the energy for
some of the configurations obtained from the London approach. The
straight horizontal line is the absolute value of the net
topological charge. \label{ExDef} }
\end{figure}

Finally, in order to further investigate the relation between
defects in the vortex configurations and the energy of the system,
we computed the total number of defects (the number of the $+$ and
$-$ topological charges) in each stable configuration obtained
within the London framework. The results are shown in
Figs.~\ref{ExDef} for $L = 110$, 111 and 112 at $H_0 = 0.1$ (left)
and $L = 230$, 232 and 234 at $H_0 = 0.2$ (right). The absolute
value of the net topological charge is depicted as a solid
horizontal line and is always equal to six as required by the
Euler theorem. The total number of defects -- which is directly
related to the number of dislocations in the configurations -- is
depicted as points connected by lines. One can notice that the
total number of defects is not a monotonic function of the London
energy of the configuration. Instead, it highly fluctuates. For
example, a configuration free of dislocations (in which only six
disclinations occur) almost always has a higher energy than, e.g.,
one with a total number of $16$ topological charges. This happens,
for example, for $L = 111$ at $H_0 = 0.1$ where such a
configuration with only six disclinations (and no dislocations)
has $\mathcal{G}_L = 0.1302066$, which is $0.5\%$ higher than the
energy of the lowest energy state, $\mathcal{G}_L = 0.12958384$
(the Voronoi contruction of the latter configuration is the $L =
111$ state depicted in Fig.~\ref{VorGL}). This indicates that the
presence, as well as the distribution, of dislocations in the
vortex configurations plays an important role in lowering the
energy of such configurations.

\section{Surface Superconductivity}
\label{secvi}
\begin{figure}[!tbp]
\centering
\begin{minipage}[t]{0.49\linewidth}
\includegraphics[width=1\textwidth]{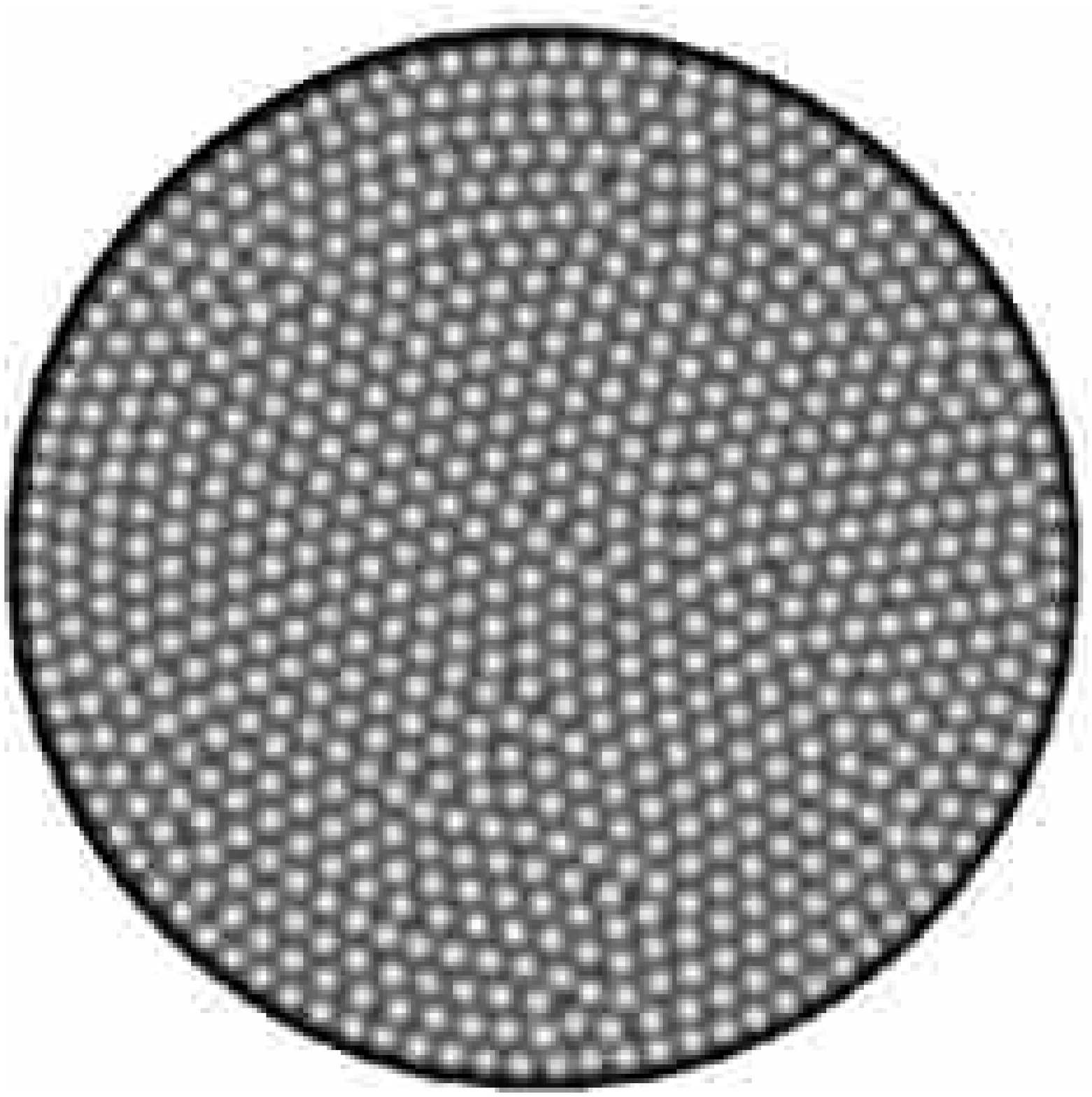}
\end{minipage}
\begin{minipage}[t]{0.49\linewidth}
\includegraphics[width=1\textwidth]{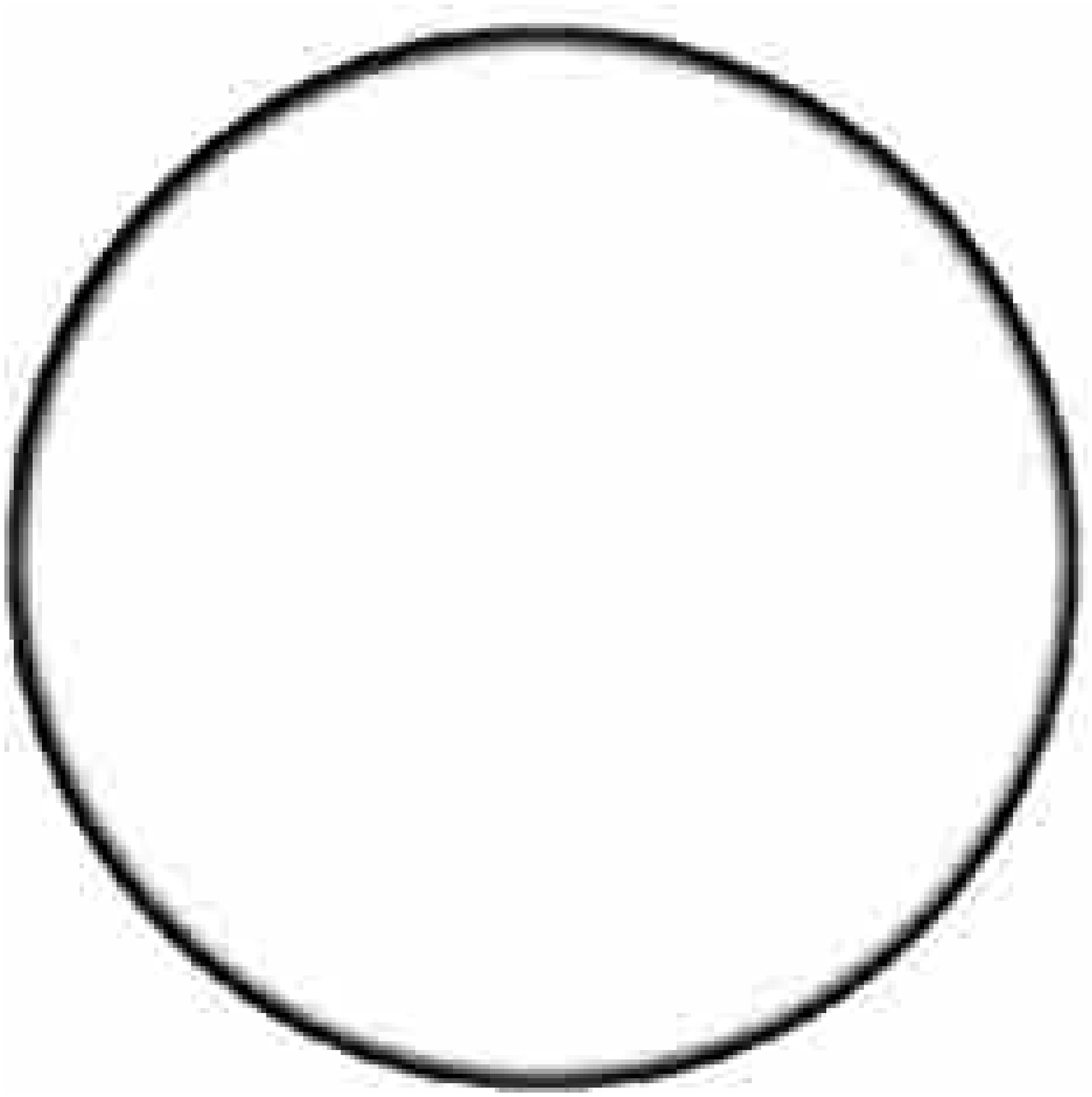}
\end{minipage}
\vspace{0.25cm} \caption{Superconducting electron density for $H =
0.6$ (left) and $H = 1.02$ (right). White to black runs from low
to high values of $|\Psi|^2$. \label{Surf1} }
\end{figure}
\begin{figure}[!tbp]
\centering
\begin{flushleft}
\includegraphics[width=0.475\textwidth]{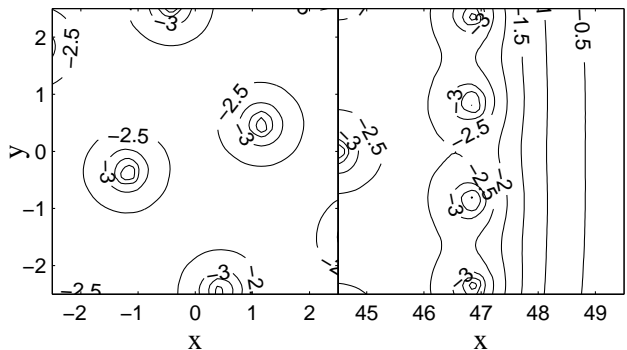}
\end{flushleft}
 \vspace{0.0cm}
\caption{Contour plots of $\log|\Psi|^2$ at the center (left) and
close to the edge of the disk for $H = 1$. \label{Surf2} }
\end{figure}

When the external magnetic field approaches $H_0 = 1$ (or $H_0 =
H_{c2}$ in not normalized units) the vorticity, $L$, becomes
large. Inside a thin layer close to the disk edge the
superconducting electron density, $|\Psi|^2$, is larger than in
the interior of the disk.~\cite{SJamesdeGennes} Such a behavior
may be understood as a result of the superposition of the
superconducting electron density depreciation close to each vortex
inside the disk, which is less strong for vortices at the surface.
This already takes place for $H_0 = 0.6$ with $L = 717$, but is
highly pronounced at $H_0 \ge 1.0$. At $H_0 = 0.6$, a multivortex
state (as was shown in previous figures and also on the left of
Fig.~\ref{Surf1}) is enclosed by this superconducting sheath.
Within this sheath $|\Psi|^2 \approx 0.75$, opposed to a maximum
of $|\Psi|^2 \approx 0.5$ between two adjacent vortices.
Nevertheless, according to the criterion adopted to characterize
the existence of a giant vortex state ($|\Psi|^2 \le 10^{-4}$ in
the region between vortices),~\cite{Baelus_PRB03} a giant vortex
state appears at $H_0 = 1.02$. In this state $|\Psi|^2 < 10^{-4}$,
except at $R - 2\xi<\rho< R$ where $0.2 \le |\Psi|^2 \le 0.45$
(cf. Fig.~\ref{Surf1} at right). At $H_0 = 1$ the maximum value of
$\left|\Psi\right|^2$ is $\sim 10^{-2}$ in the region between two
adjacent vortex cores, while $\left|\Psi\right|^2 \approx 0.55$ at
the disk edge. Such a configuration is not yet a giant vortex
state, although the multivortex state in this case is extremely
`dilute'. Possibly $H = 1$ is close to the field in which a giant
vortex state decays into a multivortex state.~\cite{ConnecSuper}
Moreover, at this magnetic field the depreciation of $|\Psi|^2$
close to the vortex cores is different whether a vortex sits in
the outermost ring or in the interior of the disk. This feature is
depicted in Fig.~\ref{Surf2}, where a contour plot of the
logarithm of the superconducting electron density is shown in the
center of the disk (at left) and close to the edge (at right).

\section{Conclusions}
\label{secviii}

We investigated the magnetic field dependence of vortex states in
thin disks with large radius. The nonlinear GL equations, as well
the London approximation were used to obtain stable vortex
configurations. Although both methods lead, for small fields, to
similar vortex configurations, the energies are different. This is
the reason for the failure of the London limit to yield the
correct ground state configuration. For low values of the
vorticity we improved the London approximation by including the
spatial variation of $|\Psi|^2$ close to the vortex cores, which
resulted in energies which were very close to those of the GL
approach.

Multivortex states were obtained for fields up to $H_0 \approx
H_{c2}$, above which a giant vortex state appears. We investigated
how the configuration of this multivortex state changes as
function of the magnetic field. At low magnetic fields ($H_0 \ll
0.1H_{c2}$) we find vortex configurations having ringlike
distribution, as expected from symmetry considerations. However as
the number of vortices increases, the vortex-vortex repulsion
starts playing a larger role and we observed the appearance of an
hexagonal lattice. The ringlike structure is replaced by an
Abrikosov lattice in the center of the disk as soon as the field
is close to $0.1H_{c2}$, when $L \sim 100$, but is preserved near
the edges. For fields larger than $0.1$ this Abrikosov lattice
becomes even more pronounced compared to the ringlike structure.

The topological defects in the vortex configurations and their
distribution were also studied. We observed two types of defects:
(wedge) disclinations and dislocations. The net topological charge
is always $-6$, as required for an hexagonal structure confined to
a circular geometry. Similar to classical particles confined in
radially symmetric potentials, we find that these topological
defects appear mostly close to the edge for $L \lesssim 230$, in
order to adjust the ringlike structure to the Abrikosov lattice.
We attribute the presence of dislocations in that region due to
the screening of disclinations. As $L$ increases further
dislocations start to be spread in the center of the disk and form
grain boundaries.

Surface superconductivity was observed at fields around and above
$0.6H_{c2}$. This surface superconductivity becomes more
pronounced as the vorticity increases, which resulted in a larger
overlap between the vortices. We also noticed that the transition
from a multivortex to a giant vortex state takes place at magnetic
fields slightly above $H_{c2}$. Just below the formation of the
giant vortex state, the superconducting electron density presents
markedly distinct spatial dependence close to the disk edge --
where the vortex structure starts to coalesce -- compared to what
is observed in the center of the disk.

\acknowledgments

This work was supported by the Flemish Science Foundation
(FWO-Vl), the ``Onderzoeksraad van de Universiteit Antwerpen''
(GOA), the ``Belgian Science Policy'', the European ESF-Vortex
Matter, and the Brazilian Science Agency CAPES. One of us (BJB)
acknowledges support from FWO-Vl and the Japanese Society for the
Promotion of Science. We acknowledge useful discussions with C. C.
de Souza Silva and M. V. ${\rm Milo\check{s}evi\acute{c}}$.


\end{document}